\documentclass[a4paper,11pt]{article}
\pdfoutput=1
\usepackage{jheppub}
\usepackage{mathtools}
\allowdisplaybreaks[4]
\usepackage[noconfig]{refstyle}
\usepackage{subfigure}
\usepackage{array,tabularx,booktabs,longtable}

\let\eqref=\relax
\newref{eq}{name={eq.~},Name={Eq.~},names={eqs.~},Names={Eqs.~},rngtxt={-},refcmd=(\ref{#1})}
\newref{app}{name={appendix~},Name={Appendix~},names={appendixes~},Names={Appendixes~}}
\newref{tab}{name={table~},Name={Table~},names={tables~},Names={Tables~}}
\newref{sec}{name={section~},Name={Section~},names={sections~},Names={Sections~}}
\newref{fig}{name={figure~},Name={Figure~},names={figures~},Names={Figures~}}
\numberwithin{equation}{section}
\newcommand{\be}{\begin{equation}}
\newcommand{\ee}{\end{equation}}
\newcommand{\field}[1]{\mathbb{#1}}
\newcommand{\oneon}[1]{\frac{1}{#1}}
\newcommand{\tr}{\mathrm{tr}}
\newcommand{\Z}{\field{Z}}
\newcommand{\CP}{\field{P}}
\newcommand{\al}{\alpha}
\newcommand{\dd}{{\rm d}}
\title{Topological Invariants and Fibration Structure of Complete Intersection Calabi-Yau Four-Folds}

\author[a]{James~Gray,}
\author[b]{Alexander~S.~Haupt}
\author[c]{and Andre~Lukas}

\affiliation[a]{Department of Physics, Robeson Hall, Virginia Tech,\\
        Blacksburg, VA 24061, U.S.A.}
\affiliation[b]{Institut f\"ur Theoretische Physik, Leibniz Universit\"at Hannover,\\
        Appelstra\ss{}e 2, 30167 Hannover, Germany}
\affiliation[c]{Rudolf Peierls Centre for Theoretical Physics, Oxford University,\\
        1 Keble Road, Oxford, OX1 3NP, U.K.}

\emailAdd{jamesgray@vt.edu}
\emailAdd{alexander.haupt@itp.uni-hannover.de}
\emailAdd{lukas@physics.ox.ac.uk}

\abstract{We investigate the mathematical properties of the class of Calabi-Yau four-folds recently found in ref.~\cite{Gray:2013mja}. This class consists of 921,497 configuration matrices which correspond to manifolds that are described as complete intersections in products of projective spaces. For each manifold in the list, we compute the full Hodge diamond as well as additional topological invariants such as Chern classes and intersection numbers. Using this data, we conclude that there are at least 36,779 topologically distinct manifolds in our list. We also study the fibration structure of these manifolds and find that 99.95 percent can be described as elliptic fibrations. In total, we find 50,114,908 elliptic fibrations, demonstrating the multitude of ways in which many manifolds are fibered. A sub-class of 26,088,498 fibrations satisfy necessary conditions for admitting sections. The complete data set can be downloaded \href{http://www-thphys.physics.ox.ac.uk/projects/CalabiYau/Cicy4folds/index.html}{{\tt here}}.}

\preprint{ITP--UH--05/14}
\arxivnumber{1405.2073}

\begin{document}
\maketitle
\flushbottom

%%%%%%%%%%%%%%%%%%%%%%%%%%%%%%%%%%%%%%%%%%%%%%%%%%%%%%%%%%%%%%%%%%%%%%%%%%%%%%%%%%%%%%%%%%%%%%%%
\section{Introduction and review of the classification of CICY four-folds}\seclabel{review}
%%%%%%%%%%%%%%%%%%%%%%%%%%%%%%%%%%%%%%%%%%%%%%%%%%%%%%%%%%%%%%%%%%%%%%%%%%%%%%%%%%%%%%%%%%%%%%%%

Calabi-Yau manifolds have played a central role in many aspects of the development of string theory, from phenomenology to formal theory. Several constructions of Calabi-Yau three-folds have seen extensive use in the literature including the hypersurfaces in toric ambient spaces~\cite{Kreuzer:2000xy,Kreuzer:2002uu} and the complete intersections in products of projective spaces~\cite{Hubsch:1986ny,Candelas:1987kf,Green:1986ck,Candelas:1987du}. Complete classes of Calabi-Yau four-folds are somewhat rarer in the literature however~\cite{Brunner:1996bu,Klemm:1996ts,Kreuzer:1997zg,Lynker:1998pb,Gray:2013mja,Anderson:2014gla}, partly due to the greater computational power which is required to exhaust these much larger data sets.

In a previous paper~\cite{Gray:2013mja}, the authors attempted to improve upon this situation by presenting a complete classification of the Calabi-Yau four-folds which can be described as complete intersection in products of projective spaces. In this paper we will expand upon various mathematical properties of these manifolds which are important for their use in physics. In order to make the present paper self-contained, however, we will begin with a brief review of the central findings of the classification presented in ref.~\cite{Gray:2013mja}.

\vspace{0.1cm}

To set up our notation, we consider a complete intersection of $K$ polynomials $p_\alpha$ in an ambient space ${\cal A}$ which is a product of $m$ projective spaces ${\cal A}= \bigotimes_{r=1}^m \CP^{n_r} $ of total dimension $K+4=\sum_{r=1}^mn_r$. We shall use indices $r,s,\ldots =1,\ldots,m$ to label the ambient projective space factors $\CP^{n_r}$ and indices $\alpha,\beta,\dots =1,\ldots ,K$ to label the defining polynomials $p_\alpha$. We can succinctly describe families of such manifolds in terms of a \emph{configuration matrix}
\be\eqlabel{conf2}
 [{\bf n}|{\bf q}] \equiv \left[\begin{array}{c|ccc}n_1 & q^1_1&\dots&q^1_K\\ \vdots & \vdots&\ddots&\vdots\\ n_m & q^m_1&\hdots&q^m_K\\\end{array}\right] ,
\ee
where the entries $q_\alpha^r$ are non-negative integers. The columns of the configuration matrix ${\bf q}_\alpha =(q_\alpha^r)_{r=1,\ldots ,m}$ denote the multi-degrees of the defining polynomials $p_\alpha$. In other words, the $\alpha$'th polynomial, $p_\alpha$, is of degree $q_\alpha^r$ in the homogeneous coordinates $x_{r,i}$ of $\CP^{n_r}$. In order to ensure that the common zero locus of these equations gives rise to a well-defined four-dimensional manifold, we demand that the $K$-form
\be
 \dd p_1 \wedge \cdots \wedge \dd p_K
\ee
is nowhere vanishing.

The family of complete intersection varieties which are described by a given configuration matrix $[{\bf n}|{\bf q}]$ is redundantly parametrized by the coefficients in the polynomials $p_\alpha$. A generic choice of theses coefficients in any particular case defines a smooth complete intersection manifold, thanks to an application of Bertini's theorem~\cite{Green:1986ck}. Many of the key properties of the manifolds depend only on the configuration matrix and not on the specific variety within the family.  Given this, in the rest of this paper, we will often not need to distinguish between the family described by the matrix $[{\bf n}|{\bf q}]$ and a specific variety therein.

A complete intersection variety of the type described above defines a Calabi-Yau manifold, denoted ${\cal X}$, if and only if the first Chern class vanishes, $c_1 ({\cal X}) = 0$. This is equivalent to the conditions
\be\eqlabel{c1zero}
 \sum_{\alpha = 1}^K q_\alpha^r = n_r + 1
\ee
on each row of the configuration matrix. In ref.~\cite{Gray:2013mja} the authors classified a complete set of such configuration matrices which describe all CICY four-folds. A priori, there is an infinite number of configuration matrices of the form~\eqref*{conf2} obeying~\eqref*{c1zero}. However, the same CICY four-fold can be described by an infinite number of \emph{different} configuration matrices. To avoid such infinite repetitions, it is possible to identify suitable equivalence relations between configurations and only keep one representative per class~\cite{Candelas:1987kf,Gray:2013mja}. A simple example of such an equivalence relation is the permutation of rows or columns. Clearly, configuration matrices related by such row or column permutations describe the same complete intersection variety since the ordering of ambient space factors and polynomials in the configuration matrix is completely arbitrary. In addition, we made use of other equivalence relations which are both practical as well as rigorously provable~\cite{Candelas:1987kf,Gray:2013mja}. This lead to a finite list classifying all topological types of CICY four-folds~\cite{Gray:2013mja}.

The complete list presented in ref.~\cite{Gray:2013mja} contains 921,497 configuration matrices in 587 different ambient spaces with a maximum matrix size of $16\times 20$. A subset of 15813 matrices corresponds to product manifolds. These fall into four types, namely $T^8$, $T^2 \times$CY$_3$, $T^4 \times K3$ and $K3 \times K3$. The Euler characteristic $\chi$ for each matrix was computed and found to be in the range $0\leq \chi \leq 2610$. All configurations with $\chi=0$ correspond to direct product manifolds and the non-zero values for the Euler characteristic were found to be in the range $288\leq \chi\leq 2610$. In total, the list contains 206 different values of $\chi$. This topological data provided a very weak lower bound on the number of inequivalent CICY four-folds. 

In the present paper we compute a number of additional topological invariants associated to these manifolds, such as Hodge numbers, Chern classes and intersection numbers. In addition to being useful in mathematical and physical applications, these results enable us to establish a significantly larger lower bound of at least 36,779 topologically distinct CICY four-folds. In the Hodge data, we find an approximate linear relation between $h^{2,2}$ and $h^{3,1}$, which is, at least to our knowledge, a mere consequence of the construction of CICY four-folds. Analyzing the pair $(h^{1,1}, h^{3,1})$, which is interchanged under mirror symmetry, we conclude that the mirror of a CICY four-fold is in most cases not itself a CICY four-fold. The only exceptions are the 153 Hodge theoretically self-mirror configurations for which $h^{1,1} = h^{3,1}$ holds. In view of potential applications for F-theory compactifications, we also study the elliptic fibration structure of CICY four-folds. We concentrate on a specific, easy-to-handle class of elliptic fibrations, which provides a rich data set of 50,114,908 elliptic fibrations distributed among 99.95 percent of the CICY four-folds. In addition, we present a classification of the different types of almost fano three-folds that occur as base manifolds and we find 26,088,498 fibrations that satisfy necessary conditions for admitting sections.

In the next section, we will describe how to compute several topological invariants associated to CICY four-folds. These will include Chern classes, Hodge data, intersection numbers of favourable divisors, and invariants constructed from the intersection numbers which do not depend upon a choice of basis for $H^{1,1}({\cal X})$. In \secref{fib}, we will describe how many of the CICY four-folds can be written as an elliptic fibration over an almost fano three-fold base. In addition we will discuss how to compute some necessary conditions for the existence of certain types of sections. In \secref{results} we will provide a cartography of the results of the computations described in \secref{topinv,fib} for the data set of the CICY four-folds computed in ref.~\cite{Gray:2013mja}. Finally, a few technical results required in the text will be provided in \appref{perms} and the format in which we present our data is explained in \appref{dataformat}.

%%%%%%%%%%%%%%%%%%%%%%%%%%%%%%%%%%%%%%%%%%%%%%%%%%
\section{Topological invariants}\seclabel{topinv}

In this section, we will describe how to compute various topological invariants of the CICY four-folds, including Chern classes, Hodge numbers and intersection numbers. These numerical characteristics are of importance in both the mathematical and physical investigation of these manifolds. Mathematically, topological invariants contain significant information about the structure of the Calabi-Yau four-fold and they help to establish which configuration matrices could describe the same four-folds. Physically, these quantities are of central importance in questions ranging from determining the number of moduli fields in four dimensions, to the structure of tadpole cancellation conditions.

\subsection{Chern classes}\seclabel{chernclasses}

For a general complete intersection manifold ${\cal X}$, not necessarily Calabi-Yau, with configuration matrix $[{\bf n}|{\bf q}]$, the first four Chern classes are given by the following expressions~\cite{Green:1986ck,Gray:2013mja}
\begin{align}
	c_1({\cal X})&=c_1^r J_r = \left[n_r+1-\sum_{\alpha=1}^Kq^r_\alpha\right]J_r \; , \eqlabel{c1} \\ 
	c_2({\cal X})&= c_2^{rs} J_r J_s = 
		\oneon{2} \left[-(n_r+1)\delta^{rs} + \sum_{\al=1}^K q_\al^r q_\al^s + c_1^r c_1^s \right] J_r J_s \; , \eqlabel{c2} \\
	c_3({\cal X})&= c_3^{rst} J_r J_s J_t = 
		\oneon{3} \left[(n_r+1)\delta^{rst} - \sum_{\al=1}^K q_\al^r q_\al^s q_\al^t + 3 c_1^r c_2^{st} - c_1^r c_1^s c_1^t \right] J_r J_s J_t \; ,\eqlabel{c3} \\
	c_4 ({\cal X}) &= c_4^{rstu} J_r J_s J_t J_u = \oneon{4}
		\left[ -(n_r+1)\delta^{rstu} + \sum_{\al=1}^K q_\al^r q_\al^s q_\al^t q_\al^u + 2 c_2^{rs} c_2^{tu} \right. \nonumber 
		\\ & \left. \qquad\qquad\qquad\qquad\qquad\qquad\qquad + 4 c_1^r c_3^{stu} - 4 c_1^r c_1^s c_2^{tu} + c_1^r c_1^s c_1^t c_1^u \vphantom{\sum_{\al=1}^K} \right] J_r J_s J_t J_u \; . \eqlabel{c4}
\end{align}
The multi-index Kronecker-symbol appearing above is defined to be $\delta^{r_1 \ldots r_n} = 1$ if $r_1 = r_2 = \ldots = r_n$ and zero otherwise. In these expressions, $J_r$ denotes the K\"ahler form of the $r$'th ambient projective space $\CP^{n_r}$, which is normalized in such a way that
\be\eqlabel{Pnorm}
 \int_{\CP^{n_r}} J_r^{n_r} = 1 \; .
\ee
For a configuration to describe a family of Calabi-Yau manifolds we need $c_1({\cal X})=0$. This leads to the Calabi-Yau constraint~\eqref*{c1zero} presented in \secref{review}. If a configuration does indeed represent a family of Calabi-Yau four-folds, the \eqref{c2,c3,c4} for the higher Chern classes simplify substantially, since all terms containing a factor of the first Chern class then vanish.

The fourth Chern class can be used to compute the Euler characteristic $\chi$ by a version of the Gauss-Bonnet formula
\be\eqlabel{Euler_c4}
 \chi ({\cal X}) = \int_{{\cal X}} c_4({\cal X}) \; .
\ee
This expression is easily evaluated using the fact that the integral of a top-form $\omega$ over ${\cal X}$ can be pulled back to an integration over the ambient space ${\cal A} = \CP_1^{n_1} \times\cdots\times \CP_m^{n_m}$. To do this we use
\be\eqlabel{mudef}
 \int_{{\cal X}} \omega = \int_{\cal A} \omega \wedge\mu_{{\cal X}} \; ,\qquad 
 \mu_{{\cal X}} \equiv \bigwedge_{\al=1}^K\left(\sum_{r=1}^mq^r_\al J_r\right) ,
\ee
and the normalizations~\eqref*{Pnorm} of the K\"ahler forms $J_r$. The $(K,K)$-form $\mu_{{\cal X}}$ is a representative of the class which is Poincar\'e dual to the homology class of the family of sub-manifolds ${\cal X} = [{\bf n}|{\bf q}]$ in the ambient space ${\cal A}$.

Given the proceeding paragraph, the explicit formula for the Euler characteristic $\chi$ of a four-fold configuration ${\cal X} = [{\bf n}|{\bf q}]$ is simply given by the following
\be\eqlabel{chi_c4}
 \chi ({\cal X}) = \left[c_4({\cal X})\wedge \mu_{{\cal X}}\right]_{\text{top}} \; .
\ee
Here, the subscript ``top'' refers to the coefficient of the volume form $J_1^{n_1} \wedge\cdots\wedge J_m^{n_m}$ of ${\cal A}$ which should be extracted from the enclosed expression.

\subsection{Hodge data}\seclabel{hodgecalc}

In terms of bundle valued cohomology, the Hodge data of the CICY four-folds may be expressed as follows
\begin{equation}\eqlabel{mrhodge1}
 H^{p,q}({\cal X})  \cong H^p({\cal X},\wedge^q {\cal TX}^*) \; .
\end{equation}
On a Calabi-Yau four-fold there are four non-trivial Hodge numbers, namely $h^{1,1}$, $h^{2,1}$, $h^{3,1}$ and $h^{2,2}$, which need to be determined for the $921,497$ configuration matrices in our data set. As such, we need an efficient procedure to calculate the cohomologies~\eqref*{mrhodge1} on a computer. We will begin our discussion with some simple relations amongst the Hodge data of a Calabi-Yau four-fold that allow us to avoid calculating some of the individual cohomologies directly. We will then describe how we compute the remaining bundle valued cohomologies directly for the geometries of interest.

On a Calabi-Yau four-fold, the Betti numbers are determined by the Hodge numbers as follows
\begin{equation}
\begin{split}
 &b^0=b^8=1 \;,\qquad b^1=b^7 =0 \;,\qquad b^2=b^6=h^{1,1} \;,\\&b^3=b^5=2 h^{2,1} \;,\qquad b^4=2 h^{3,1}+h^{2,2}+2 \; .
\end{split}
\end{equation}
The Euler characteristic of the manifold, a topological invariant for which we reviewed a simple formula in \secref{chernclasses}, can likewise be expressed in terms of the Betti numbers. We obtain the following expression relating the Euler and Hodge numbers
\begin{equation}
 \chi= \sum_{q=0}^8 (-1)^q b^q = 4 + 2 h^{1,1}-4 h^{2,1} + 2 h^{3,1} + h^{2,2} \; . \eqlabel{chi}
\end{equation}
Thus, one of the Hodge numbers is determined by the others and the Euler characteristic, enabling us to avoid calculating one of the cohomologies~\eqref*{mrhodge1} explicitly.

Another simplification of this type can be achieved by considering the indices $\chi_q = \chi({\cal X}, \wedge^q {\cal TX}^*)$. From the index theorem, we have
\begin{equation}\eqlabel{indextheorem}
 \chi_q = \sum_{p=0}^4 (-1)^p h^{p,q}({\cal X}) = \int_{\cal X} \text{ch}(\wedge^q {\cal TX}^*) \wedge \text{Td}({\cal TX}) \;.
\end{equation}
The splitting principle formulae, $c({\cal TX}) = \sum_i(1+x_i)$, $\text{ch}({\cal TX})= \sum_i e^{x_i}$ and
\begin{equation} 
 \text{ch}(\wedge^q {\cal TX}^*) \wedge \text{Td}({\cal TX}) = \sum_{i_1 > \ldots > i_q} e^{-x_{i_1}} \ldots e^{-x_{i_q}} \prod_j \frac{x_j}{1-e^{-x_j}},
\end{equation}
together with the Calabi-Yau condition $c_1({\cal X})=0$ of the four-fold, can be used to show that the indices~\eqref*{indextheorem} take the following form
\begin{align}\eqlabel{eulerc2c2}
 \chi_0 &= 2 =\frac{1}{720} \int_{\cal X} (3 c_2^2 -c_4) \; ,\\
 \chi_1 &= -h^{1,1}+h^{2,1}-h^{3,1} = \frac{1}{180} \int_{\cal X} (3 c_2^2 - 31 c_4) \; ,\\ 
 \chi_2 &= -2 h^{2,1}+ h^{2,2} = \frac{1}{360} \int_{\cal X}(9c_2^2  + 237 c_4) \; ,\\
 \chi_3 &= \chi_1 \;\;,\;\; \chi_4 = \chi_0 \; .
\end{align}
From this we get one additional non-trivial relation: $22 \chi_0 - 4 \chi_1 - \chi_2 =0$, which gives us
\begin{equation}\eqlabel{rel_hodge_numbers}
 -4 h^{1,1} + 2 h^{2,1} - 4 h^{3,1} + h^{2,2} =44 \; .
\end{equation}

\vspace{0.1cm}

The direct computation of the remaining two Hodge numbers is performed using the theory of spectral sequences~\cite{Hartshorne:1977,Griffiths:1978,Distler:1987ee,Hubsch:1992nu}\footnote{See also ref.~\cite{Anderson:2008ex} for a nice introduction to these kinds of computations.}. We will make use of two key short exact sequences
\begin{equation}\eqlabel{euler}
 0 \to {\cal O}_{{\cal X}}^m \to {\cal R}\to {\cal T}_{\cal A}|_{\cal X} \to 0\;,\qquad
 0 \to {\cal T}_{\cal X} \to {\cal T}_{\cal A} |_{\cal X} \to {\cal N} \to 0 \; ,
\end{equation} 
referred to as the Euler and adjunction sequence, respectively. The normal bundle ${\cal N}$ and the bundle ${\cal R}$ can both be written as sums of line bundles and are explicitly given by\footnote{In this paper, we are using the following standard notation for line bundles on products of projective spaces and CICYs. The line bundle ${\cal O}_{\cal A}( k^r)$ on ${\cal A}$ is that whose first Chern class is given by $c_1({\cal O}_{\cal A}(k^r))= k^r J_r$. The line bundle ${\cal O}_{\cal X}(k^r)$ is the restriction of ${\cal O}_{\cal A}( k^r)$ to the Calabi-Yau four-fold.}
\begin{equation}
{\cal N}= \bigoplus_{a=1}^K {\cal O}_{\cal X}({\bf q}_a) \;,\qquad {\cal R}=\bigoplus_{r=1}^m {\cal O}_{\cal X}({\bf e}_i)^{\oplus (n_r+1)}\; ,
\end{equation} 
where ${\bf e}_i$ are the standard unit vectors. The long exact sequences associated to these two short exact sequences can be written in the form
\begin{equation*}
 \begin{array}{c|ccccc|cccccc}
  &{\cal O}_{{\cal X}}^m&\to& {\cal R}&\to& {\cal T}_{\cal A}|_{\cal X}&&{\cal T}_{\cal X}&\to&{\cal T}_{\cal A} |_{\cal X}&\to&{\cal N} \\\hline
 H^0({\cal X},\cdot)& \mathbb{C}^m&&H^0({\cal X},{\cal R})&&H^0({\cal X}, {\cal T}_{\cal A}|_{\cal X})&&0&&
 H^0({\cal X},{\cal T}_{\cal A} |_{\cal X})&&H^0({\cal X},{\cal N})\\
 H^1({\cal X},\cdot)& 0&&H^1({\cal X},{\cal R})&&H^1({\cal X}, {\cal T}_{\cal A}|_{\cal X})&&H^{3,1}({\cal X})&&
 H^1({\cal X},{\cal T}_{\cal A} |_{\cal X})&&H^1({\cal X},{\cal N})\\
 H^2({\cal X},\cdot)& 0&&H^2({\cal X},{\cal R})&&H^2({\cal X}, {\cal T}_{\cal A}|_{\cal X})&&H^{2,1}({\cal X})&&
 H^2({\cal X},{\cal T}_{\cal A} |_{\cal X})&&H^2({\cal X},{\cal N})\\
 H^3({\cal X},\cdot)& 0&&H^3({\cal X},{\cal R})&&H^3({\cal X}, {\cal T}_{\cal A}|_{\cal X})&&H^{1,1}({\cal X})&&
 H^3({\cal X},{\cal T}_{\cal A} |_{\cal X})&&H^3({\cal X},{\cal N})\\
 H^4({\cal X},\cdot)& \mathbb{C}^m&&H^4({\cal X},{\cal R})&&H^4({\cal X}, {\cal T}_{\cal A}|_{\cal X})&&0&&
 H^4({\cal X},{\cal T}_{\cal A} |_{\cal X})&&H^4({\cal X},{\cal N})
\end{array}  
\end{equation*}
From the first of these long exact sequences we learn that 
\begin{equation}
\begin{array}{lllllll}
 H^0({\cal X},{\cal T}{\cal A}|_{\cal X})&\cong& \frac{H^0({\cal X},{\cal R})}{\mathbb{C}^m}&\quad\quad&
 H^1({\cal X},{\cal T}{\cal A}|_{\cal X})&\cong&H^1({\cal X},{\cal R})\\
 H^2({\cal X},{\cal T}{\cal A}|_{\cal X})&\cong&H^2({\cal X},{\cal R})&\quad\quad&
 H^4({\cal X},{\cal T}{\cal A}|_{\cal X})&\cong& H^4({\cal X},{\cal N})\; 
\end{array}
\end{equation}
and 
\begin{equation}
 H^3({\cal X},{\cal T}{\cal A}|_{\cal X})\cong H^3({\cal X},{\cal R})+{\rm Ker}(\mathbb{C}^m\rightarrow H^4({\cal X},{\cal R}))\eqlabel{H3}\; .
\end{equation} 
Combining these results with the second long exact sequence gives
\begin{align}
 H^{3,1}({\cal X})&\cong \frac{H^0({\cal X},{\cal N})}{H^0({\cal X},{\cal R})/\mathbb{C}^m}\oplus {\rm Ker}(H^1({\cal X},{\cal R})\rightarrow H^1({\cal X},{\cal N}))\label{h31}\\
 H^{2,1}({\cal X})&\cong {\rm Coker}(H^1({\cal X},{\cal R})\rightarrow H^1({\cal X},{\cal N}))\oplus {\rm Ker}(H^2({\cal X},{\cal R})\rightarrow H^2({\cal X},{\cal N}))\\
 H^{1,1}({\cal X})&\cong {\rm Coker}(H^2({\cal X},{\cal R})\rightarrow H^2({\cal X},{\cal N}))\oplus {\rm Ker}(H^3({\cal X},{\cal T}{\cal A}|_{\cal X})\rightarrow H^3({\cal X},{\cal N}))
\end{align}
for the desired Hodge cohomologies. The main observation from these results is that the Hodge numbers can be computed entirely from the bundle cohomology of the line bundle sums ${\cal N}$ and ${\cal R}$ on ${\cal X}$. 

A particularly interesting sub-class consists of those CICY four-folds which are favour\-able. We call a CICY four-fold favourable if its complete second cohomology descends from the second cohomology of the ambient space, so that $H^{1,1}({\cal X})\cong\mathbb{C}^m$, where $\mathbb{C}^m$ is the space which appears in \eqref{H3}. A sufficient (although slightly too strong) set of conditions for this to be the case is
\begin{equation}
 H^2({\cal X},{\cal N})=H^3({\cal X},{\cal N})=H^3({\cal X},{\cal R})=H^4({\cal X},{\cal R})=0\; . \eqlabel{favour}
\end{equation} 
Provided these conditions hold the Hodge numbers for favourable CICYs satisfy
\begin{equation}\eqlabel{hfav}
\begin{split}
 h^{1,1}({\cal X})&=m\; ,\\ h^{3,1}({\cal X})-h^{2,1}({\cal X})&=m-h^0({\cal X},{\cal R})+h^1({\cal X},{\cal R})-h^2({\cal X},{\cal R})+h^0({\cal X},{\cal N})-h^1({\cal X},{\cal N})\; . 
\end{split}
\end{equation}
Together with the Euler number constraint, \eqref{chi}, this fixes three Hodge numbers in terms of the line bundle cohomology of ${\cal N}$ and ${\cal R}$ without the need to compute ranks of maps.\footnote{It seems that, in this situation, the additional constraint, \eqref{rel_hodge_numbers}, fixes the fourth Hodge number. However, it turns out that this constraint is usually automatically implied by \eqref{hfav} and \eqref{chi}.}

To complete the Hodge number calculation, we need to be able to compute line bundle cohomology on CICYs and, in general, determine the ranks of maps between such cohomologies. The first step in this direction is to relate a line bundle ${\cal L}_{\cal A}$ on the ambient space ${\cal A}$ to its restriction ${\cal L}={\cal L}_{\cal A}|_{\cal X}$ onto ${\cal X}$ by the Koszul resolution, a long exact sequence given by~\cite{Hartshorne:1977,Griffiths:1978,Matsumura:1986,Hubsch:1992nu}
\begin{equation}
  0\rightarrow \wedge^K{\cal N}_{\cal A}^*\otimes {\cal L}_{\cal A}\rightarrow \dots \rightarrow \wedge^2{\cal N}_{\cal A}\otimes {\cal L}_{\cal A}^*\rightarrow {\cal N}_{\cal A}^*\otimes {\cal L}_{\cal A}\rightarrow {\cal L}_{\cal A}\rightarrow {\cal L}\rightarrow 0\; .
\end{equation}  
This long exact sequence can be broken up into short exact sequences each of which have associated long exact sequences in cohomology or, alternatively, we can study the spectral sequence associated to the Koszul resolution. Either way, this allows for the computation of line bundle cohomology on the CICY ${\cal X}$ in terms of ambient space line bundle cohomology. Line bundle cohomology on a single projective space is described by a theorem due to Bott, Borel and Weil, see for example~\cite{Hubsch:1992nu}. To obtain the cohomology for line bundles on our ambient space, which are products of projective spaces, we can simply apply a version of K\"unneth's formula to the result for single projective spaces. In this way, we can develop an algorithm to compute line bundle cohomology on CICYs and, combined with the above results, this allows for a computation of Hodge numbers.

\subsection{Intersection numbers and distinguishing invariants}\seclabel{intnums}

Besides Chern classes, Hodge numbers and the Euler characteristics, there are a number of additional invariants which can be used to distinguish different topological types of CICY four-folds. We will focus, in particular, on those that are easily computable from the configuration matrix.

We begin by introducing a basis $\{\nu^i\}$ of $H^6({\cal X})$, dual to the integral basis, $\{J_i\}$, of $H^2({\cal X})$ such that
\be
 \int_{\cal X} J_i \wedge \nu^j = \delta_i^j
\ee
as usual. Of course, the products $J_i \wedge J_j \wedge J_k$ can be written as linear combinations of the basis $\{\nu^i\}$ and it follows easily that
\be
 J_i \wedge J_j \wedge J_k = d_{ijkl} \nu^l \; , \qquad 
 d_{ijkl} = \int_{\cal X} J_i \wedge J_j \wedge J_k \wedge J_l \; ,
\ee
where $d_{ijkl}$ are the quadruple intersection numbers.

The products $J_i \wedge J_j$ for $i \leq j$ can be thought of as elements of $H^4({\cal X})$, but it is not clear that they are linearly independent. Consider a linear relation $\lambda^{ij} J_i \wedge J_j = 0$ among them. Then it follows that $d_{ijkl} \lambda^{kl} = 0$. In other words, if $d_{ijkl} \lambda^{kl} = 0$ does not have non-trivial solutions $\lambda^{kl}$ or, equivalently, if the matrix $d_{(ij)(kl)}$ has maximal rank, then the forms $J_i \wedge J_j$ for $i \leq j$ are linearly independent.

Now, consider the total Chern class expanded as
\be
 c({\cal X}) = \cdots + C_2^{ij} J_i \wedge J_j + C_3^{ijk} J_i \wedge J_j \wedge J_k + \cdots = \cdots + C_2^{ij} J_i \wedge J_j + c_{3,i} \nu^i + \cdots \; ,
\ee
where we define $c_{3,i} = d_{ijkl} C_3^{jkl}$, $c_{2,ij} = d_{ijkl} C_2^{kl}$ and so on. Clearly, one may form an invariant in the following way
\be\eqlabel{c2inv}
 I = \int_{\cal X} c_2({\cal X}) \wedge c_2({\cal X}) = c_{2,ij} C_2^{ij} = \tr(c_2 C_2) \; .
\ee
However, due to~\eqref*{eulerc2c2}, this invariant carries the same information as the Euler characteristic in the case of a Calabi-Yau four-fold. The problem with other contractions which involve $c_2$ or $C_2$ is that they may represent the second Chern class in a redundant way, since the forms $J_i \wedge J_j$ may not be linearly independent. So, in the absence of other assumptions, \eqref{c2inv} seems to be the only further invariant which involves Chern classes. 

Let us, for the moment, assume that the forms $J_i \wedge J_j$ for $i \leq j$ are linearly independent, a condition which can be explicitly checked from the intersection numbers in any given case. Then we have the following additional invariants
\begin{align}
 I_p &= c_3 C_2 (c_2 C_2)^p c_3 \; , \\
 \tilde{I}_q &= \tr((c_2 C_2)^q) \; ,
\end{align}
for $p \geq 0$ and $q \geq 1$. Unfortunately, with the exception of the small configurations at the beginning of the list, the forms $J_i \wedge J_j$ practically always turn out to be linearly dependent and hence the above invariants $I_p$ and $\tilde{I}_q$ are of little practical use.

Next, we turn to invariants extracted solely from the quadruple intersection numbers. We will follow the logic of ref.~\cite{Green:1988fr} and generalise their results to CICY four-folds. We begin by defining the intersection form
\be
 \Lambda(K_1, K_2, K_3, K_4) = \int_{\cal X} K_1 \wedge K_2 \wedge K_3 \wedge K_4 \; ,
\ee
where $K_1, \ldots, K_4$ represent classes in $H^2({\cal X}, \Z)$. In terms of this form, the quadruple intersection numbers are of course given by
\be\eqlabel{dLambda}
 d_{ijkl} = \Lambda(J_i, J_j, J_k, J_l) \; .
\ee
The next step is to define the following sets
\begin{align}
 S_1 &= \{\Lambda(K_1, K_2, K_3, K_4) \,|\, K_a \in H^2({\cal X},\Z) \} \; , \\
 S_2 &= \{\Lambda(K_1, K_2, K_3, K_3) \,|\, K_a \in H^2({\cal X},\Z) \} \; , \\
 S_3 &= \{\Lambda(K_1, K_2, K_2, K_2) \,|\, K_a \in H^2({\cal X},\Z) \} \; , \\
 S_4 &= \{\Lambda(K_1, K_1, K_1, K_1) \,|\, K_1 \in H^2({\cal X},\Z) \} \; ,
\end{align}
and ${\cal I}_p = \mathrm{gcd}(S_p)$. The virtue of the above sets is that they are not only topologically invariant but, unlike the intersection numbers themselves, they are also basis-independent and hence the ${\cal I}_p$ are genuine invariants. This invariance is due to the sets being defined by scanning over the entire integral lattice spanned by $K_a \in H^2({\cal X},\Z)$. However, computing the intersection form on all elements of $H^2({\cal X},\Z)$ is not practical. Instead, a simplification is achieved by expanding $K_a = n_a^i J_i$ with integer coefficients $n_a^i$, which leads to
\be\eqlabel{expandLambdad}
 \Lambda(K_1, K_2, K_3, K_4) = d_{ijkl} n_1^i n_2^j n_3^k n_4^l \; .
\ee
If two or more arguments in $\Lambda(K_1, K_2, K_3, K_4)$ are identical, the quadruple sums on the right hand side can be decomposed into smaller building blocks according to
\begin{align}
 &d_{ijkl} n_1^i n_1^j n_2^k n_3^l = d_{iijk} (n_1^i)^2 n_2^j n_3^k + 2 \sum_{i<j} d_{ijkl} n_1^i n_1^j n_2^k n_3^l  \ , \eqlabel{d1123_expansion}\\
 &d_{ijkl} n_1^i n_1^j n_1^k n_2^l = d_{iiij} (n_1^i)^3 n_2^j + 3 \sum_{i<j} \left[ d_{ijjk} n_1^i (n_1^j)^2 n_2^k + d_{iijk} (n_1^i)^2 n_1^j n_2^k \right] \nonumber\\
            &\qquad\qquad\qquad\,+ 6 \sum_{i<j<k} d_{ijkl} n_1^i n_1^j n_1^k n_2^l \; , \eqlabel{d1112_expansion}\\
 &d_{ijkl} n_1^i n_1^j n_1^k n_1^l = d_{iiii} (n_1^i)^4 + 6 \sum_{i<j} d_{iijj} (n_1^i)^2 (n_1^j)^2  + 4 \sum_{i<j} \left[ d_{ijjj} n_1^i (n_1^j)^3 + d_{iiij} (n_1^i)^3 n_1^j \right] \nonumber\\
            &\qquad\qquad\qquad\,+ 12 \sum_{i<j<k} \left[ d_{ijkk} n_1^i n_1^j (n_1^k)^2 + d_{ijjk} n_1^i (n_1^j)^2 n_1^k + d_{iijk} (n_1^i)^2 n_1^j n_1^k \right] \nonumber\\
            &\qquad\qquad\qquad\,+ 24 \sum_{i<j<k<l} d_{ijkl} n_1^i n_1^j n_1^k n_1^l \; . \eqlabel{d1111_expansion}
\end{align}
Given this, we define, in addition,\footnote{The sign choices in the definitions of $\tilde{S}_3$ and $\tilde{S}_4$ arise because we want to compare these sets to $S_3$ and $S_4$ which scan over the entire integral lattice. This includes, in particular, those elements which have an expansion of the form $K = \pm J_1 \pm J_2 \pm \ldots$ in terms of the basis $\{J_i\}$. From \eqref{d1112_expansion,d1111_expansion}, we see that all possible relative signs appear in the ordered sums that involve ambiguities in the grouping of indices. Thus, we need to include all possible relative signs in order for the entire lattice to be scanned by $\tilde{S}_3$ and $\tilde{S}_4$.}
\begin{align}
 \tilde{S}_1 &= \{ d_{ijkl} \,|\, i,j,k,l = 1,\ldots,h^{1,1}({\cal X}) \} \; , \\
 \tilde{S}_2 &= \{d_{iijk}| i,j,k = 1,\ldots,h^{1,1}({\cal X})\} \cup \{ 2 d_{ijkl} |  i,j,k,l = 1,\ldots,h^{1,1}({\cal X})\} \; , \\
 \tilde{S}_3 &= \{d_{iiij}| i,j = 1,\ldots,h^{1,1}({\cal X})\} \cup \{ 3 (d_{iijk} \pm d_{ijjk}) |  i,j,k = 1,\ldots,h^{1,1}({\cal X})\}) \nonumber\\ &\cup \{ 6 d_{ijkl} |  i,j,k,l = 1,\ldots,h^{1,1}({\cal X})\} \; , \\
 \tilde{S}_4 &= \{d_{iiii}| i = 1,\ldots,h^{1,1}({\cal X})\} \cup \{6d_{iijj}| i,j = 1,\ldots,h^{1,1}({\cal X})\} \nonumber\\ &\cup \{4 (d_{iiij} \pm d_{ijjj})| i,j = 1,\ldots,h^{1,1}({\cal X})\} 
     \nonumber\\ &\cup \{ 12 (d_{ijkk} \pm d_{ijjk} \pm d_{iijk}) |  i,j,k = 1,\ldots,h^{1,1}({\cal X})\}) \nonumber\\ &\cup \{ 24 d_{ijkl} |  i,j,k,l = 1,\ldots,h^{1,1}({\cal X})\} \; .
\end{align}
From \eqref{dLambda}, it follows that $\tilde{S}_p \subset S_p$. Therefore, a common divisor of $S_p$ is also a common divisor of $\tilde{S}_p$. Conversely, a common divisor of $\tilde{S}_p$ divides all
$\Lambda(K_1, K_2, K_3, K_4)$ owing to the expansion~\eqref*{expandLambdad} and the fact that the $n_a^i$ are integers. Altogether, this shows that $S_p$ and $\tilde{S}_p$ have equal sets of common divisors and hence, in particular
\be
 {\cal I}_p = \mathrm{gcd}(S_p) = \mathrm{gcd}(\tilde{S}_p) \; .
\ee
In practice these invariants can, of course, only be explicitly calculated for favourable configurations where we know all of the intersection numbers. These are roughly half of the CICYs in our data set.

Another invariant we consider is the signature of the intersection matrix. Denote by $G$ the matrix $d_{(ij)(kl)}$ with $i \leq j$ and $k \leq j$, where we combine the first and last two indices each into a single index. Then $G$ transforms under a change of basis as $G \to P^T G P$ with certain general linear matrices $P$. The eigenvalues of $G$ are of course not invariant under such a transformation but, by Sylvester's law of inertia, the numbers of positive and negative eigenvalues are. Hence, the two invariants obtained in this way are the number of positive and negative eigenvalues of $G$. Of course, the actual computation of theses invariants is also restricted to the favourable configurations.

%%%%%%%%%%%%%%%%%%%%%%%%%%%%%%%%%%%%%%%%%%%%%%%%%%
\section{Fibration structure}\seclabel{fib}

\subsection{A class of elliptic fibrations}\seclabel{fib_class}

We would like to enumerate and present the different ways in which the CICY four-folds discussed in \secref{review} and ref.~\cite{Gray:2013mja} can be written as an elliptic fibration over a three dimensional base. Finding every rewriting of a Calabi-Yau four-fold as an elliptic fibration turns out to be a formidable task, especially in non-favourable cases where not all of the divisors in ${\cal X}$ descend from divisors on the ambient space projective factors. Nevertheless, there exist specific types of elliptic fibration which can be simply distinguished from the structure of the configuration matrices~\eqref*{conf2} themselves. These bear some similarity to methods of identifying fibrations in other Calabi-Yau four-fold constructions~\cite{Kreuzer:1997zg}. We have performed an exhaustive classification of these readily accessible fibration structures within our data set.

Consider a configuration matrix which can, by row and column permutations, be put in the following form
\begin{equation}\eqlabel{mrfib}
{\cal X}= \left[\begin{array}{c|cc}  {\cal A}_1 & 0 & {\cal F} \\
{\cal A}_2 & {\cal B} & {\cal T} \end{array}\right] .
\end{equation}
Here ${\cal A}_1$ and ${\cal A}_2$ are two products of $N_1$ and $m-N_1$ projective spaces respectively, while ${\cal F}, {\cal B}$ and ${\cal T}$ are sub-block matrices. If ${\cal X}$ is a Calabi-Yau four-fold then all of the rows, in particular the first $N_1$, obey the condition~\eqref*{c1zero}. Thus, if the components of ${\cal F}$ are denoted $f_{\hat{\alpha}}^{\hat{r}}$ where $\hat{r}=1,\ldots,N_1$ and $\hat{\alpha}=1,\ldots, \hat{K}$, then we have $\sum_{\hat{\alpha}=1}^{\hat{K}} f_{\hat{\alpha}}^{\hat{r}} = n_{\hat{r}}+1$ and $[{\cal A}_1| {\cal F}]$ is also Calabi-Yau. In examples where $\sum_{\hat{r}=1}^{N_1} n_{\hat{r}} - \hat{K} =1$ this Calabi-Yau is a one-fold, that is, $[{\cal A}_1| {\cal F}]$ is an elliptic curve. In such a situation, the configuration matrix~\eqref*{mrfib} describes an elliptic fibration over the almost fano three-fold base $\left[ {\cal A}_2| {\cal B} \right]$ (here, ``almost fano'' is shorthand for a three-fold configuration whose anticanonical bundle is almost-ample~\cite{Hubsch:1992nu}) with the fibre being described by the matrix $[{\cal A}_1| {\cal F}]$. The twisting of the fibre over the base is encoded in the matrix ${\cal T}$. We shall refer to an elliptic fibration of the form~\eqref*{mrfib} as  an ``obvious elliptically fibration'' or ``OEF'' for short.

A given configuration matrix may admit many different OEFs of the form~\eqref*{mrfib}. In enumerating the inequivalent fibrations of this type, we face redundancy issues similar to those encountered in the compilation of the CICY four-fold list itself. It is clear that two different configuration matrices in the form~\eqref*{mrfib} can describe the same OEF, for example, if they are related by permutations of rows and columns which do not mix up the block form of the matrix. In general, the redundancy between fibrations could be due to any of the types of identities between configurations we have discussed in ref.~\cite{Gray:2013mja}. Redundancies can be removed from the description of sets of possible elliptic fibrations using very similar observations to those made in the compilation of the CICY four-fold list~\cite{Gray:2013mja}, and results such as those in \appref{perms}. We remove all of the redundancies that are enumerated in Section 4.III of reference~\cite{Gray:2013mja}, as well as row and column permutations which do not mix up the fibre and base structure described in \eqref{mrfib}. Even once such redundancies are removed, we will see that the CICY four-folds generically admit many OEFs, especially for manifolds with larger Picard number. It should be noted that not all such elliptic fibrations of a manifold can be manifest in the configuration matrix simultaneously. Some of the rows comprising the fibre in one elliptic fibration may also appear in the fibre description of an inequivalent OEF. As such, while the configuration matrix can always be put in the form \eqref{mrfib} for any single fibration, further nesting of such structure can not be assumed in the case of multiple fibrations. We also remind the reader that each configuration matrix describes an entire complex structure moduli space of configurations. Thus two `equivalent' fibrations could differ from one another for different choices of complex structure. The statement is simply that, for two equivalent fibrations, given a choice of complex structure of the first, there is a choice of complex structure of the second such that the two fibrations are identical. For more general recent results on the subject of elliptic fibrations of Calabi-Yau four-folds see ref.~\cite{Kollar:2012pv}.

Testing whether a given matrix is of the form~\eqref*{mrfib} can be straightforwardly implemented on a computer, as can the redundancy removal described above. We have analysed the list of CICY four-folds in this way and the results of this investigation will be presented in \secref{results_fibs}.

\subsection{The class of sections} \seclabel{sec}

Much of the physics literature that has been developed for describing F-theory compactifications on Calabi-Yau four-folds relies not only on the existence of an elliptic fibration, but also on that fibration admitting a section. We thus wish to study a class of sections of the OEFs discussed in the previous subsection. 

If a section exists for a given fibration, it constitutes a divisor of the Calabi-Yau four-fold itself. In the description of our manifolds we have one set of divisors over which we have particularly good calculation control - those that descend from hyperplanes in the ambient space. Divisors which descend in this way from ${\cal A}$ to ${\cal X}$ are frequently referred to as ``favourable" in the literature. For computational ease we will restrict our attention to sections which correspond to favourable divisors, referring to these as ``favourable sections". As we will see this will provide us with a very large set of examples with which to work, and thus this choice is not overly restrictive.

Deciding which of the OEFs we shall enumerate admit a favourable section is somewhat beyond the computational scope of the current paper. Instead, we will check a condition which is necessary if a fibration is to admit a section which is a generic representative of a favourable divisor class. We first define a form on the base $[{\cal A}_2|{\cal B}]$ as follows
\begin{equation}
\mu_{\text{points}}=\bigwedge_{\check{r}=N_1+1}^m  J_{\check{r}}^{n_{\check{r}}} \; .
\end{equation}
The form $\mu_{\text{points}}$ is dual to a fixed number of points in the usual way
\begin{equation}
\int_{[{\cal A}_2|{\cal B}]} \mu_{\rm{points}} = \#_{\rm points} \; .
\end{equation}
We then write a form, $S = a^r J_r$, which is dual to a general favourable divisor class. We wish to find coefficients $a^r$ for which this divisor class could contain the putative section. To do this we demand that the divisor class $S$ intersects the form dual to $\#_{\rm points}$ fibres, described by the pullback under the fibration map $\pi$ of $\mu_{\rm  points}$,  $\#_{\rm points}$ times (once for each fibre)
\begin{equation} \eqlabel{secc}
\int_{\cal X} \pi^* \mu_{\rm{points}} \wedge S =  \#_{\rm  points} \; .
\end{equation}
If there is a solution, $a^r$, to~\eqref*{secc} then the intersection numbers of ${\cal X}$ satisfy the necessary condition for a generic element of the divisor class dual to $S=A^rJ_r$ to be a section. If not, no such section can exist. We emphasize that even in the case of a positive result one has to be careful. In order to prove the existence of a section, one would need to show that there is a representative of the relevant divisor class which is nowhere vertical over the base.

%%%%%%%%%%%%%%%%%%%%%%%%%%%%%%%%%%%%%%%%%%%%%%%%%%
\section{Results}\seclabel{results}

We have applied the methods discussed in this paper to the list of 921,497 CICY four-fold configuration matrices presented in ref.~\cite{Gray:2013mja} in order to further explore their mathematical properties. In this section, we present the main results of this analysis. The complete data, which is the output of several computer programs running in parallel on a computer cluster for several months, can be downloaded from~\cite{cicylist4} in a format that is described in \appref{dataformat}.

\subsection{Cartography of properties: Hodge data and distinguishing invariants}

Using the techniques presented in \secref{hodgecalc}, we have computed all Hodge numbers of all CICY four-folds. We have excluded from this analysis the 15,813 block-diagonal configuration matrices since they correspond to product manifolds, which generally have more non-zero entries in their Hodge diamond than an indecomposable four-fold. However, the Hodge numbers in these cases follow from those of their lower-dimensional constituents and K\"unneth's formula.

\begin{figure}[!t]\centering
\includegraphics[width=0.6\textwidth]{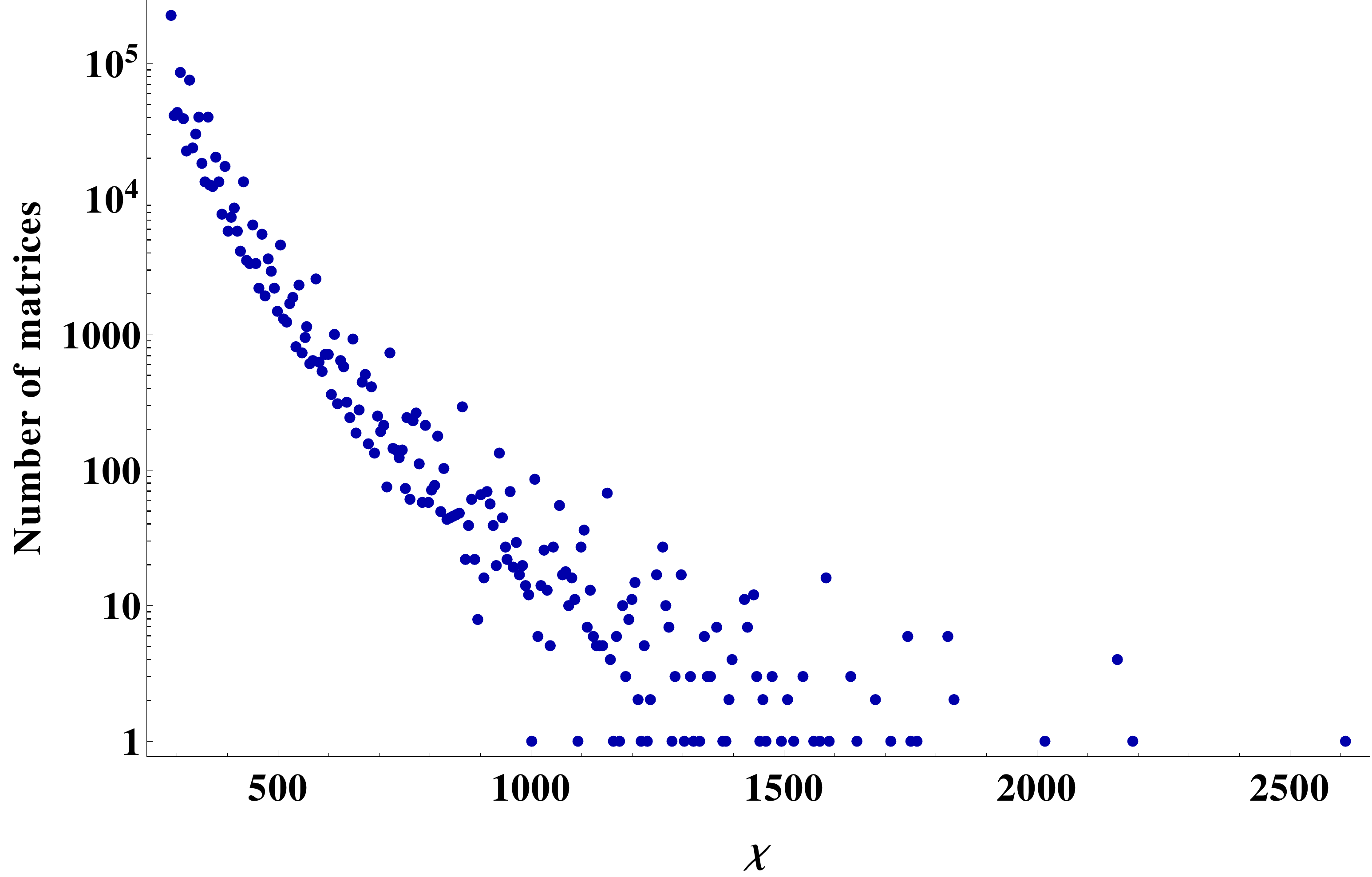}
\caption{Distribution of the Euler characteristic $\chi$ in the CICY four-fold list (excluding product manifolds), as a logarithmic plot. The values lie in the range $288 \leq \chi \leq 2610$.}
\figlabel{eulerhisto}
\end{figure}
For the remaining $\text{921,497} - \text{15,813} = \text{905,684}$ CICY four-folds, we find the following mean values for the Euler characteristic $\chi$ and the Hodge numbers $h^{p,q}$
\begin{equation}\eqlabel{hodge_mean_values}
\begin{aligned}
 \langle\chi\rangle = 341^{2610}_{288} \; , \quad
 &\langle h^{1,1} \rangle = 10.1^{24}_{1} \; , 
 &\langle h^{2,1} \rangle = 0.817^{33}_{0} \; , \\
 &\langle h^{3,1} \rangle = 39.6^{426}_{20} \; , 
 &\langle h^{2,2} \rangle = 241^{1752}_{204} \; ,
\end{aligned}
\end{equation}
where the superscripts and subscripts respectively denote the maximal and minimal values that occur. For further details, we refer to the logarithmic plots of the distribution of the Euler and Hodge numbers which can be found in \figref{eulerhisto,hodgehisto} respectively.
\begin{figure}[!t]\centering
\includegraphics[width=0.85\textwidth]{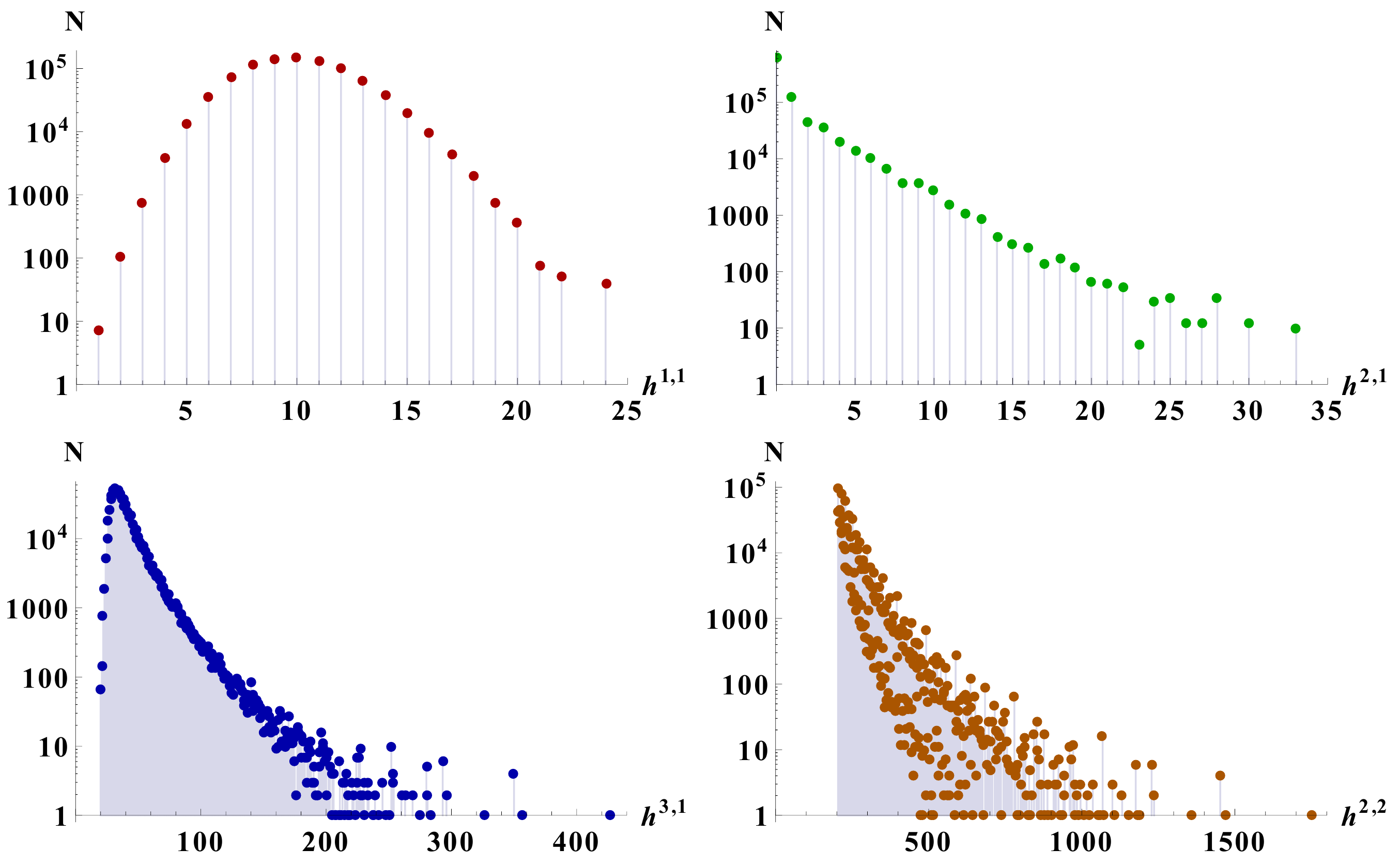}
\caption{Logarithmic plots of the abundance of Hodge numbers in the CICY four-fold list (excluding product manifolds). Here, $N$ is the number of times a given value of the Hodge number appears in the CICY four-fold list.}
\figlabel{hodgehisto}
\end{figure}
The four-dimensional space spanned by the Hodge numbers is depicted in \figref{hodgehisto2dsections}, where the six canonical two-dimensional subspaces are shown.
\begin{figure}[!t]\centering
\subfigure{\includegraphics[width=0.328\textwidth]{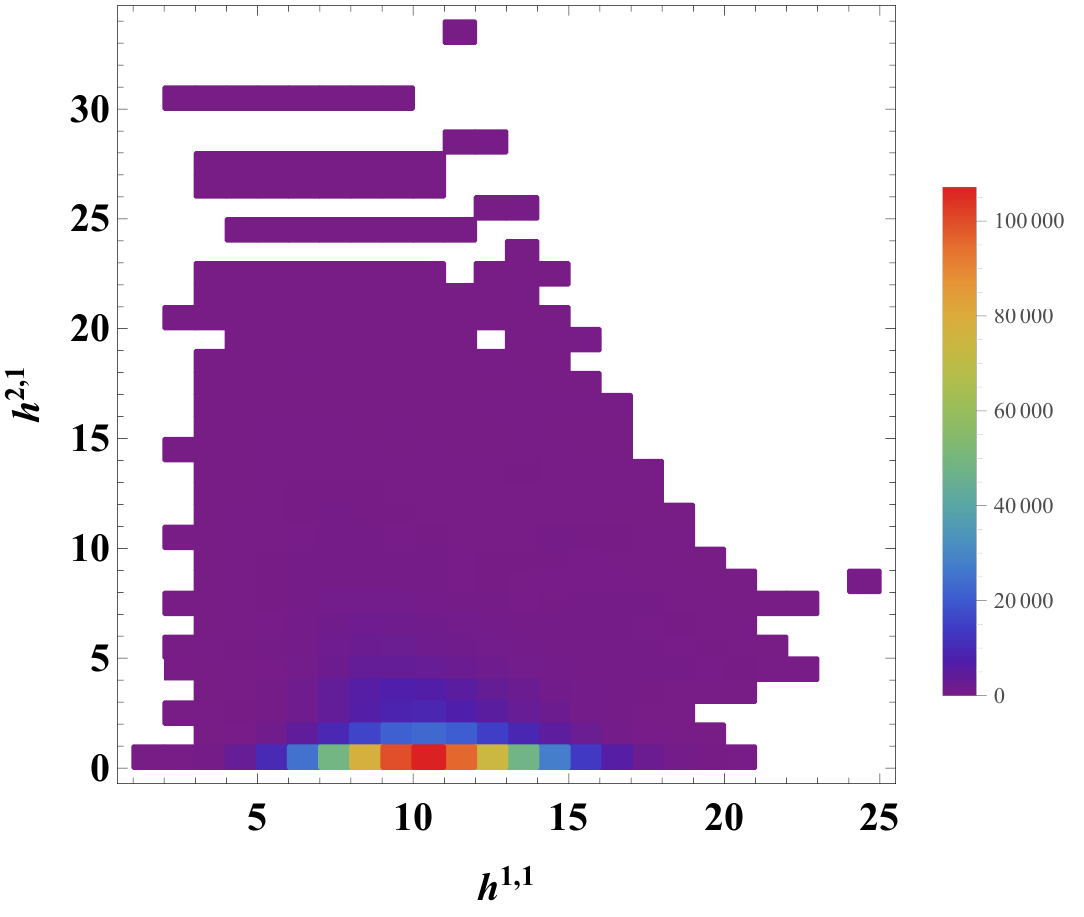}}
\subfigure{\includegraphics[width=0.328\textwidth]{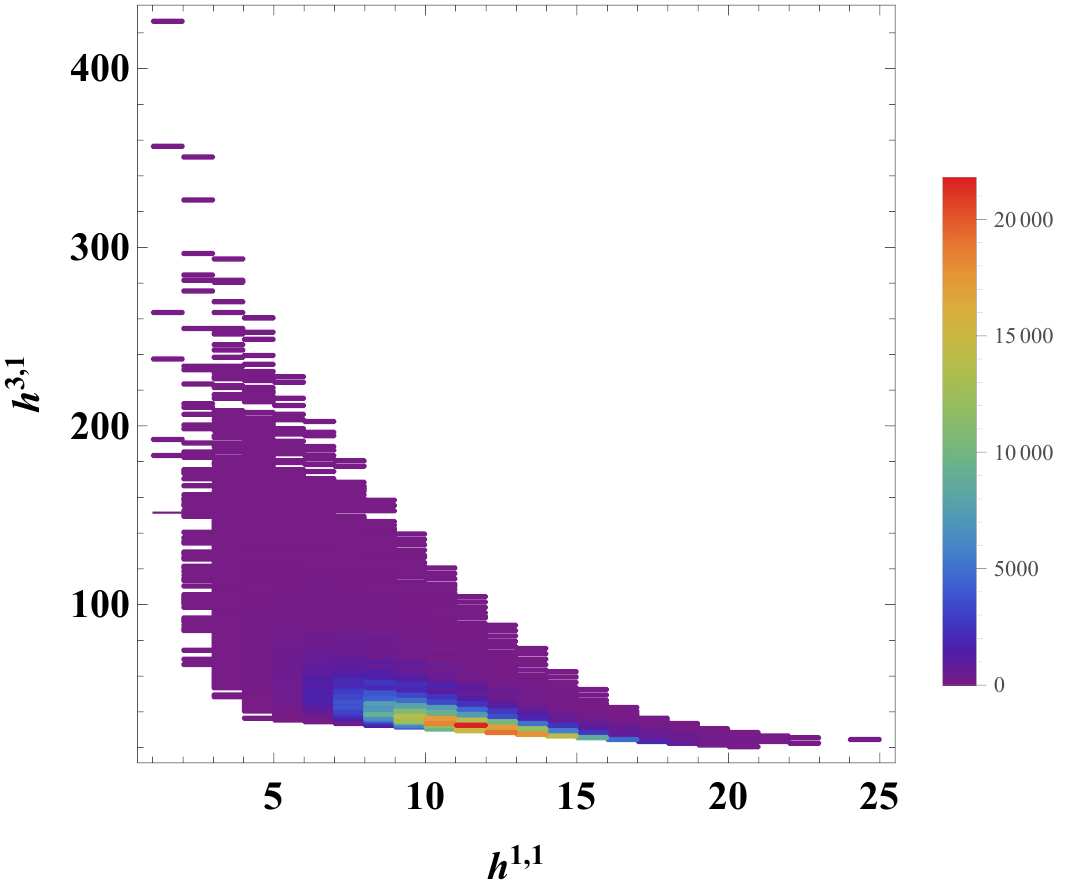}}
\subfigure{\includegraphics[width=0.328\textwidth]{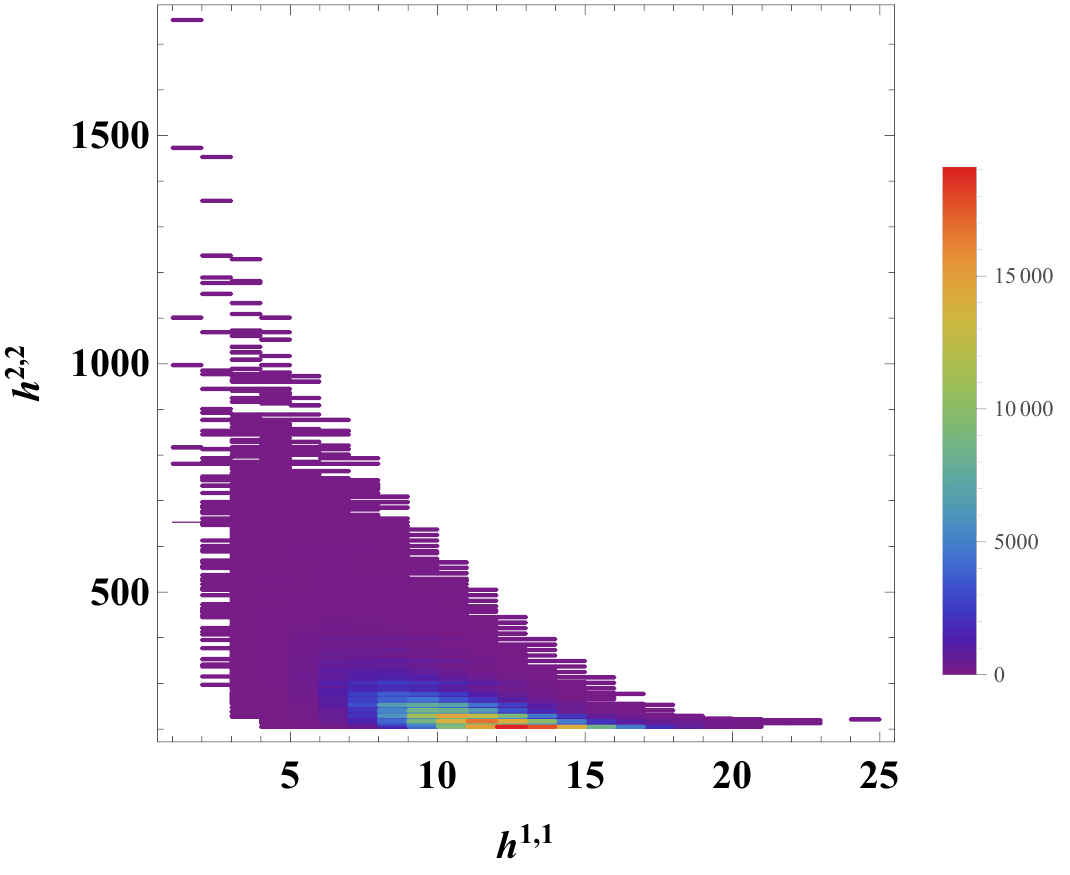}}\\
\subfigure{\includegraphics[width=0.328\textwidth]{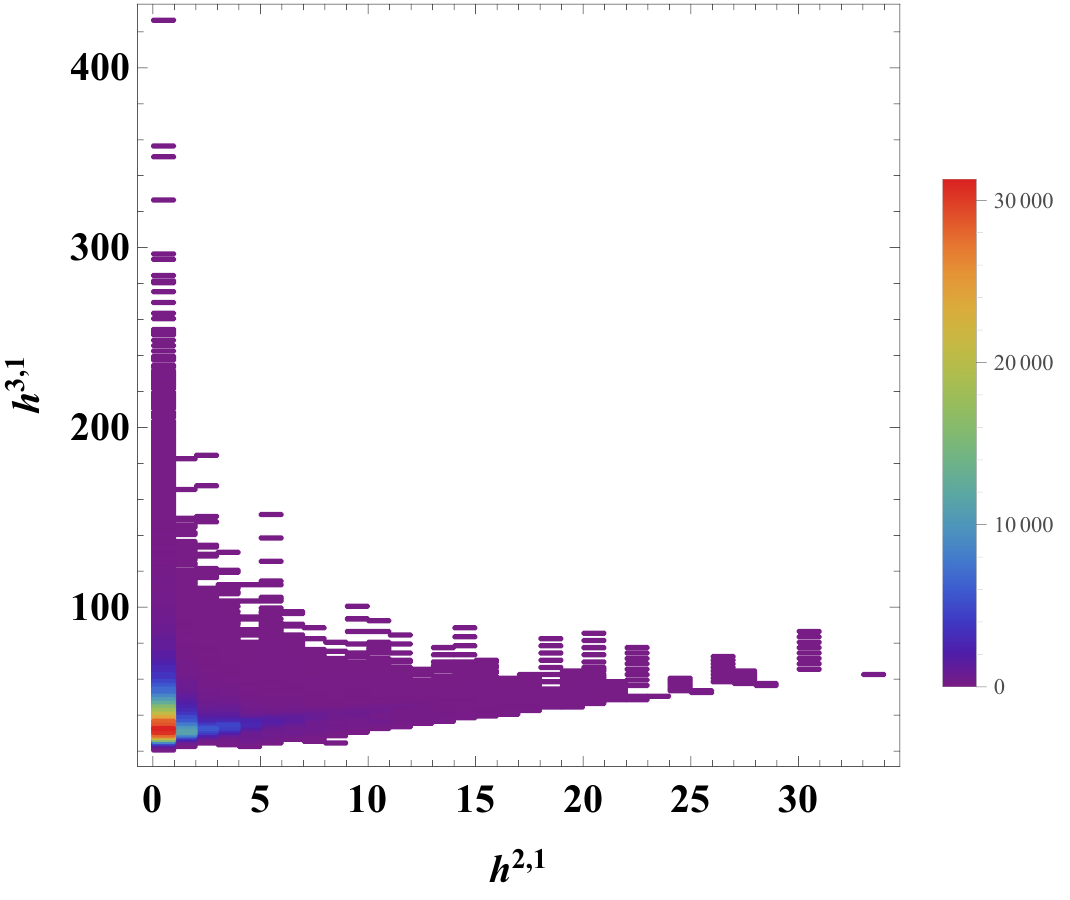}}
\subfigure{\includegraphics[width=0.328\textwidth]{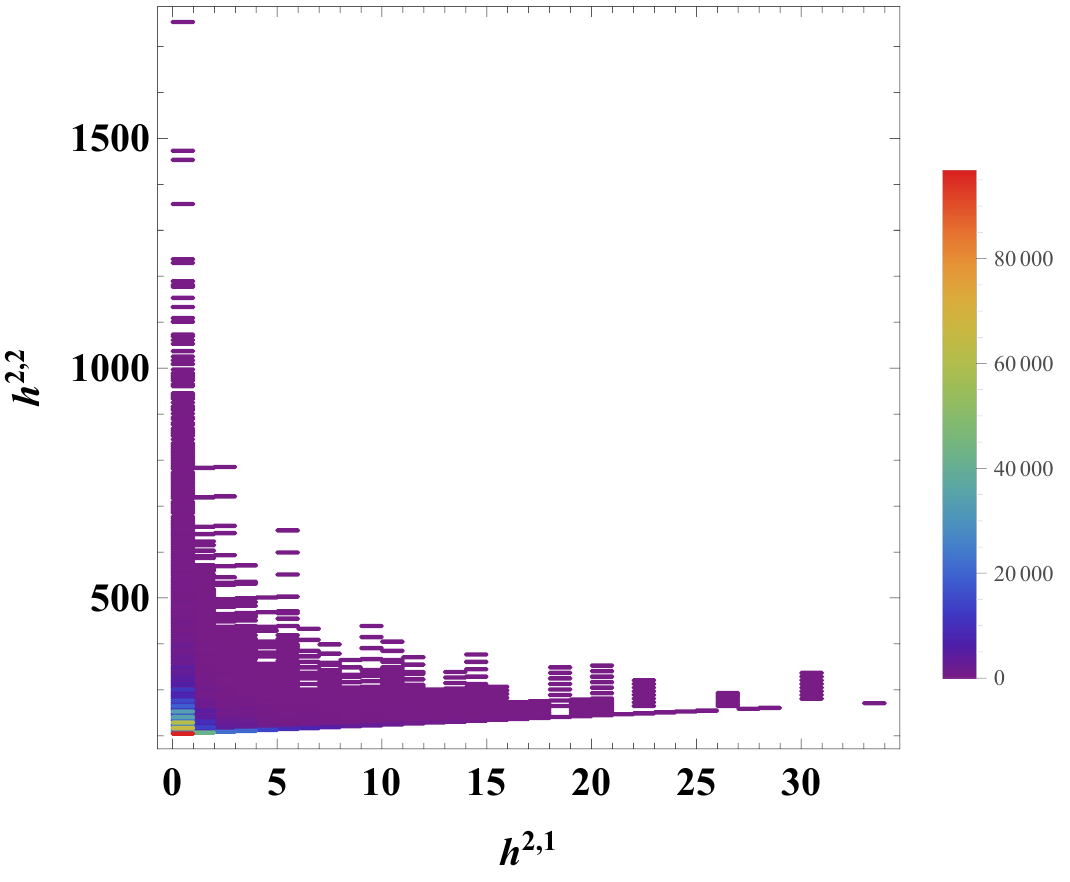}}
\subfigure{\includegraphics[width=0.328\textwidth]{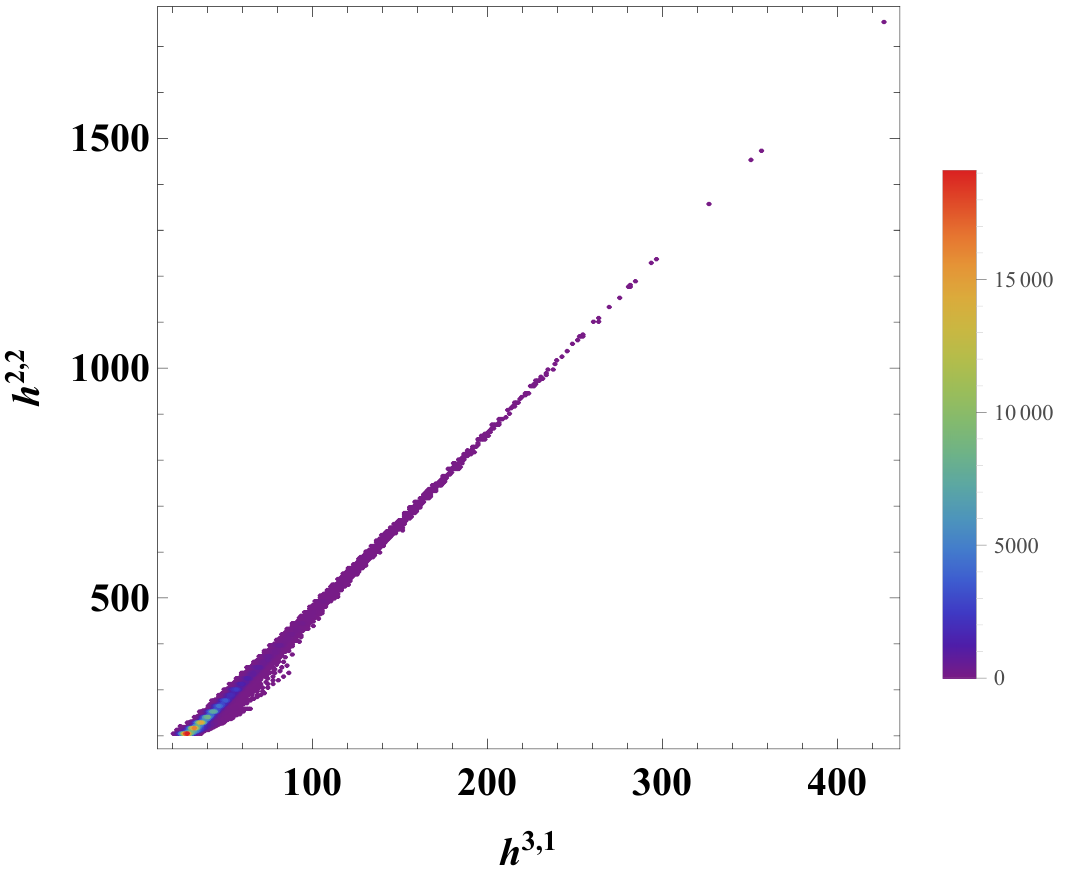}}
\caption{The canonical two-dimensional sections of the space spanned by the Hodge numbers. The colouring encodes the abundance with which the particular combination of Hodge numbers occurs in the CICY four-fold list (excluding product manifolds).}
\figlabel{hodgehisto2dsections}
\end{figure}

There is one noteworthy peculiarity evident in one of the plots.\phantomsection\label{rel_h31_h22_origin} Upon inspection of the bottom right graph in \figref{hodgehisto2dsections}, one discovers a weak correlation between the values of $h^{3,1}$ and $h^{2,2}$. The origin of this correlation is purely empirical and can be explained by considering~\eqref*{rel_hodge_numbers} rewritten as
\be
 h^{2,2} = 4 h^{3,1} + ( 44 + 4 h^{1,1} - 2 h^{2,1} ) \; .
\ee
Comparing with~\eqref*{hodge_mean_values} shows that on average $h^{2,2}$ and $4 h^{3,1}$ are an order of magnitude larger than the expression in parenthesis. To a good approximation, we may thus replace $h^{1,1}$ and $h^{2,1}$ by their mean values to obtain a linear relationship between $h^{3,1}$ and $h^{2,2}$
\be\eqlabel{linrel}
 h^{2,2} \approx 4 h^{3,1} + ( 44 + 4 \langle h^{1,1} \rangle - 2 \langle h^{2,1} \rangle ) = 4 h^{3,1} + 82.8 \; .
\ee
The graph of this approximate relationship overlaid with the exact density histogram is plotted in \figref{rel_h31_h22} showing good agreement between the linear curve and the density distribution.
\begin{figure}[!t]\centering
\includegraphics[width=0.50\textwidth]{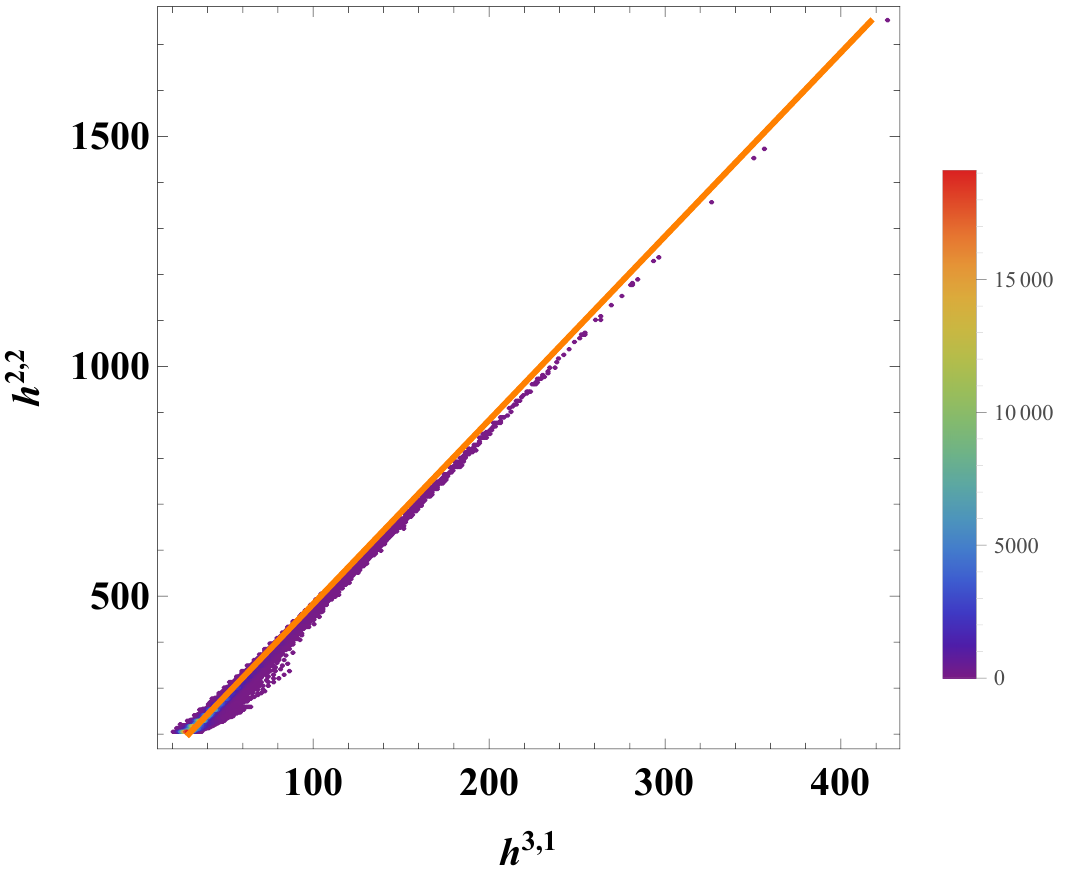}
\caption{Density histogram of the pair $(h^{3,1}, h^{2,2})$ in the CICY four-fold list (excluding product manifolds) overlaid with the linear equation $h^{2,2} \approx 4 h^{3,1} + 82.8$ (orange curve). The purely empirical origin of this apparent linear relation between $h^{3,1}$ and $h^{2,2}$ is explained on page \pageref{rel_h31_h22_origin}.}
\figlabel{rel_h31_h22}
\end{figure}
A somewhat similar relation is also known to hold for a different class of explicitly constructed Calabi-Yau four-folds~\cite{Lynker:1998pb}. It is important to stress however that the approximate linear correlation~\eqref*{linrel} is, as far as we know, merely an artefact of the construction of CICY four-folds.

Under mirror symmetry, the two Hodge numbers $h^{1,1}$ and $h^{3,1}$ are interchanged~\cite{Batyrev:1993}. In order to illustrate the situation of mirror symmetry for CICY four-folds, we show a mirror plot in \figref{mirrorplot} -- that is, a plot of $(h^{1,1} + h^{3,1})$ against $(h^{1,1} - h^{3,1})$.
\begin{figure}[!t]\centering
\includegraphics[width=0.60\textwidth]{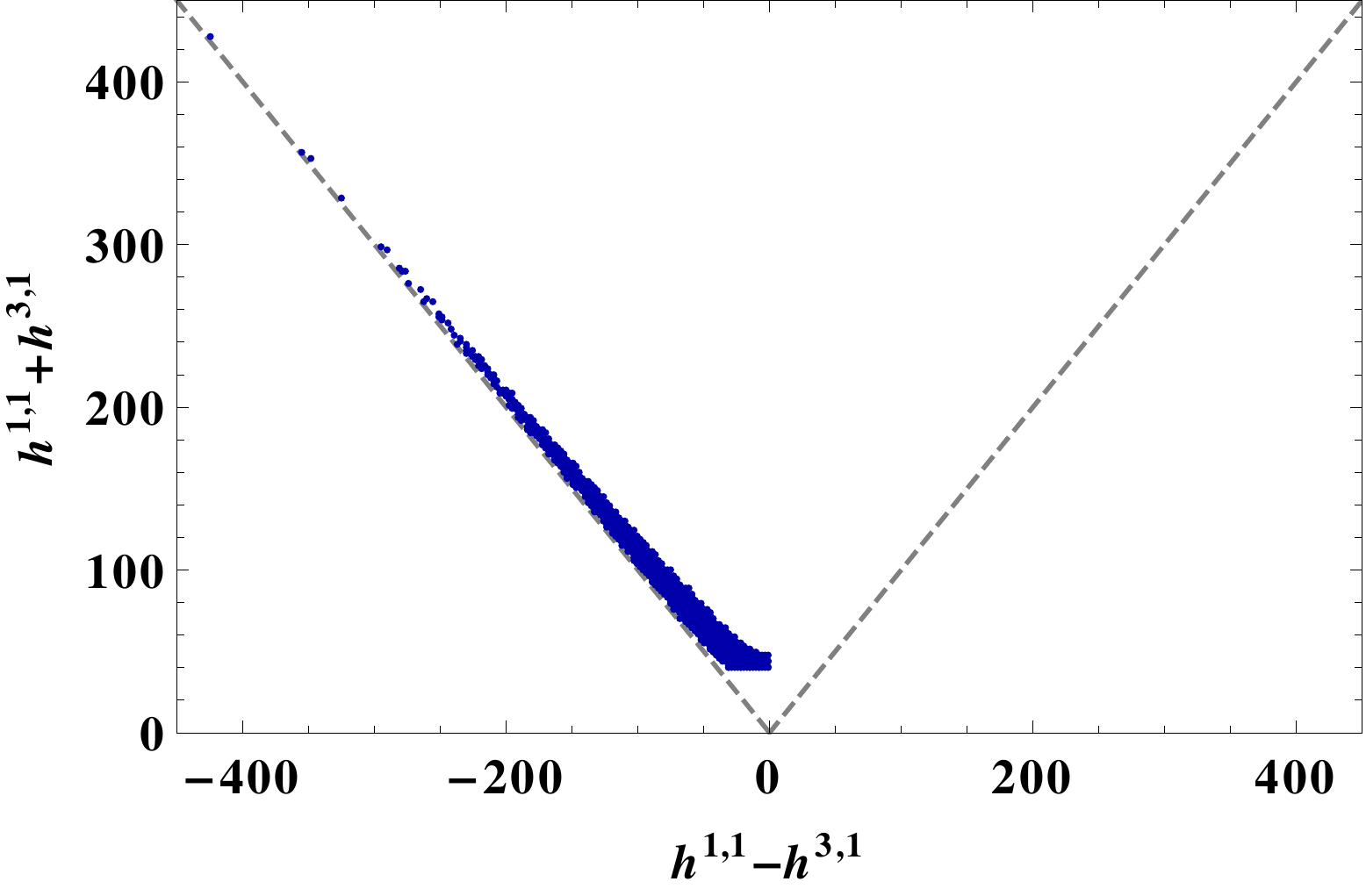}
\caption{A plot of $(h^{1,1} + h^{3,1})$ against $(h^{1,1} - h^{3,1})$. The dashed lines bound the region $h^{1,1} \geq 0$, $h^{3,1} \geq 0$.}
\figlabel{mirrorplot}
\end{figure}
From the highly asymmetrical plot, we can conclude that the mirror of a CICY four-fold is in most cases not itself a CICY four-fold, with the notable exception of 153 configurations with $h^{1,1} = h^{3,1}$. This situation is very similar to the case of CICY three-folds and, as was noted for example in ref.~\cite{Candelas:2007ac}, it is a consequence of the fact that CICYs are a rather special sub-class among \emph{all} Calabi-Yau four-folds. Indeed, for more general constructions the mirror plot typically becomes more symmetrical~\cite{Lynker:1998pb}.

Clearly, it is desirable to know how many topologically distinct manifolds there are in the list of CICY four-folds. We have thus also computed the topological invariants discussed in \secref{intnums}. Taking all of these into account, we find in total 36,779 different sets of topological invariants. This number serves as a new lower bound for the number of topologically distinct manifolds in the list of CICY four-folds. It improves the lower bound given in ref.~\cite{Gray:2013mja} by an order of magnitude. It may well be possible to raise this lower bound further by considering additional topological invariants, such as the ones studied in ref.~\cite{Bizet:2014uua}.

\subsection{The list of elliptic fibrations and their favourable sections}\seclabel{results_fibs}

We have performed an exhaustive computer scan to find all OEF structures of the type described in \secref{fib_class} among the list of CICY four-folds. The resulting data set contains 50,114,908 elliptic fibrations\footnote{We would like to remark that the numbers quoted in this section are to be regarded with some care. Despite modding out by some types of equivalences such as row and column permutations, we expect the presence of residual equivalences among the elliptic fibrations, just as with the four-fold configuration matrices themselves.} distributed among 921,020 CICY four-folds. The remaining 477 CICY four-folds cannot be brought into the OEF form~\eqref*{mrfib}. 
For the rest of this section, we exclude the 15,813 product manifolds. This reduces the number of elliptic fibrations by 648,660 to 49,466,248. On average a CICY four-fold thus admits 54.6 OEFs and the range of the number of OEFs per configuration is 0 - 354. A logarithmic plot of the distribution of the elliptic fibration abundance is shown in \figref{fibhisto}.
\begin{figure}[!t]\centering
\includegraphics[width=0.60\textwidth]{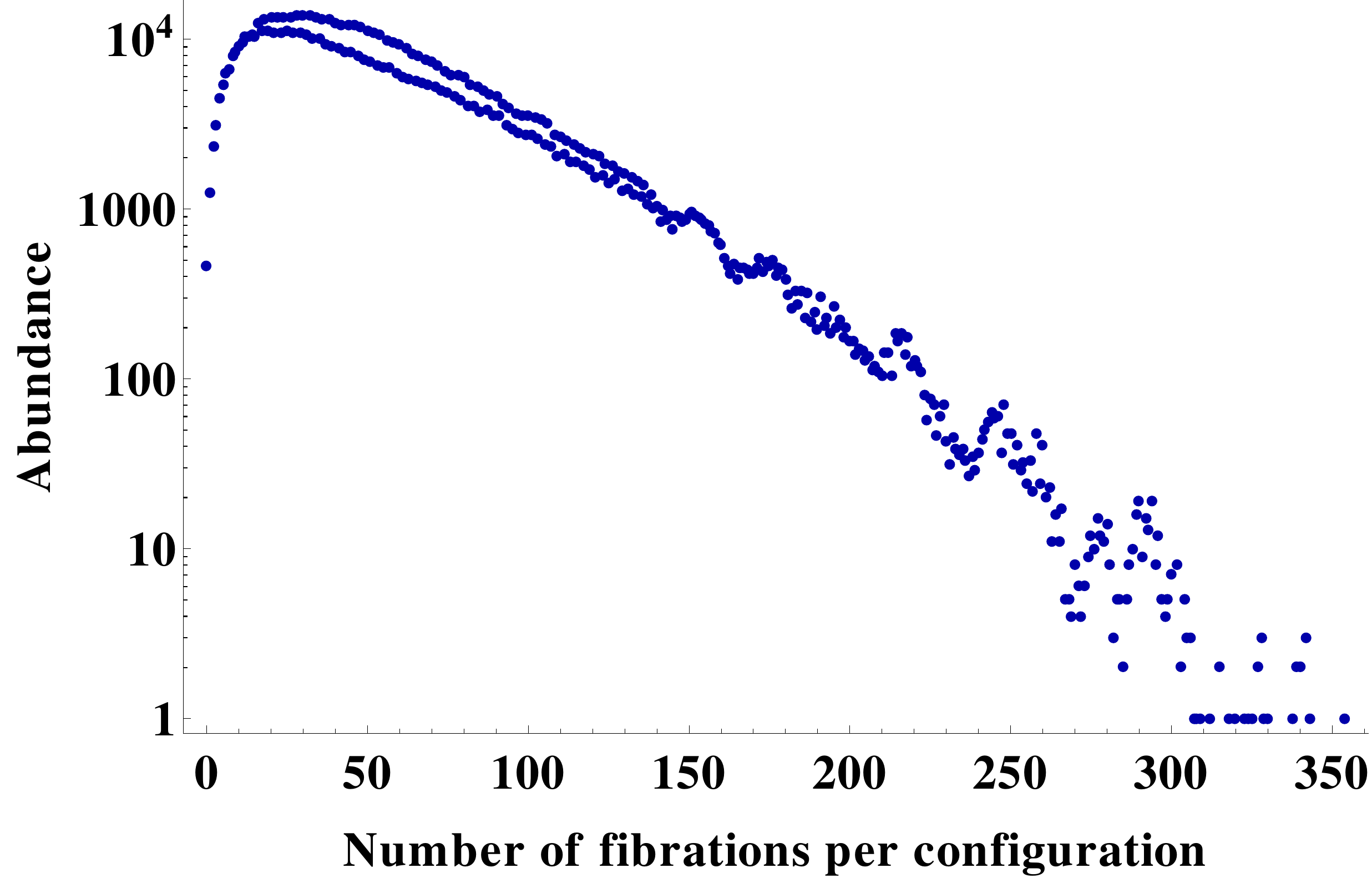}
\caption{Distribution of elliptic fibration abundance in the CICY four-fold list (excluding product manifolds). The values lie in the range 0 - 354. We find 49,466,248 OEFs in total and on average each CICY four-fold configuration is elliptically fibered in 54.6 different ways.}
\figlabel{fibhisto}
\end{figure}

It should be noted that every configuration matrix in the list with $h^{1,1} > 12$ admits at least one OEF. This ubiquity of elliptic fibrations at high Picard number echoes a structure that was found in ref.~\cite{Taylor:2012dr}, for the Kreuzer-Skarke classification of Calabi-Yau three-folds constructed as hypersurfaces in toric ambient spaces~\cite{Kreuzer:2000xy,Kreuzer:2002uu}.

There are two different types of fibre configurations in our list. The first type is given by block-diagonal fibre configurations, such as, for example
\be
 \left[\begin{array}{c|cc} 1 & 2 & 0 \\ 2 & 0 & 3 \end{array}\right] \; .
\ee
This configuration describes two points in $\CP^1$ times a torus $[2|3]$. The fibres of a total of 2,149,222 OEFs are block-diagonal. The remaining 47,317,026 non block-diagonal fibre configurations matrices describe irreducible tori. These fibrations can degenerate over special loci in the base. It should be noted that some $99.4\%$ of the fibre descriptions contain linear constraints in the coordinates of a single projective space. However, such linear constraints cannot be removed (by replacing the relevant $\CP^n$ with $\CP^{n-1}$) as different redundant descriptions of the fibre can be twisted over the base of the OEF in inequivalent ways.

It is also of interest to analyse the base manifolds that occur in our list. There are three main types of base manifolds, namely products of projective spaces, almost fano complete intersections in products of projective spaces and $\CP^1$ times almost del Pezzo complete intersections in products of projective spaces. In \tabref{bases} we further sub-divide the three main types and present a complete classification of the base manifolds that occur in our list.
\begin{table}[!t]\centering
\setlength\extrarowheight{4pt}
\begin{tabular}{c@{\hspace{3.7mm}}c@{\hspace{3.7mm}}c@{\hspace{3.7mm}}c} \toprule
Type & $\#$ & $\chi$ & Example configurations \\[4pt] \toprule 
$\CP^3$ & 562,342 & 4 & --- \\[4pt] \hline
$\CP^1 \times \CP^2$ & 9,745,787 & 6 & --- \\[4pt] \hline
$(\CP^1)^3$ & 10,030,442 & 8 & --- \\[4pt] \hline
\begin{tabular}{@{}c@{}}almost \\ fano ${\cal B}_3$ \end{tabular} & 6,252,997 & \begin{tabular}{@{}c@{}}$\{-60, -50,$ \\ $-48, -46\}$ \\ $\cup \{2\, n | -21$ \\ $\leq n\leq 12\}$\end{tabular} & \begin{tabular}{@{}cccc@{}} $\left[\begin{smallmatrix} 1 \\ 4 \end{smallmatrix} \big| \begin{smallmatrix} 0 & 2 \\ 3 & 1 \end{smallmatrix}\right]_{\text{\tiny -30}}$, & $\left[\begin{smallmatrix} 1 \\ 3 \end{smallmatrix} \big| \begin{smallmatrix} 2 \\ 2 \end{smallmatrix}\right]_{\text{\tiny 0}}$, & $\left[\begin{smallmatrix} 2 \\ 2 \end{smallmatrix} \big| \begin{smallmatrix} 1 \\ 3 \end{smallmatrix}\right]_{\text{\tiny 6}}$, & {\footnotesize $\ldots$}, \end{tabular} \\[4pt] \hline
fano ${\cal B}_3^\prime$ & 6,995,514 & \begin{tabular}{@{}c@{}}$\{-56, -36,$ \\ $-24, -14,$ \\ $-12, -6,$ \\ $-4, 0, 2, 4,$ \\ $6, 8, 10\}$\end{tabular} & \begin{tabular}{@{}cccc@{}} {\footnotesize $[4|4]_{_{-56}}$}, & {\footnotesize $[5|2\;\,2]_{_{0}}$}, & {\footnotesize $[4|2]_{_{4}}$}, & $\left[\begin{smallmatrix} 2 \\ 3 \end{smallmatrix} \big| \begin{smallmatrix} 1 & 1 \\ 1 & 2 \end{smallmatrix}\right]_{\text{\tiny -4}}$, \\ $\left[\begin{smallmatrix} 3 \\ 3 \end{smallmatrix} \big| \begin{smallmatrix} 1 & 1 & 1 \\ 1 & 1 & 1 \end{smallmatrix}\right]_{\text{\tiny 0}}$, & $\left[\begin{smallmatrix} 2 \\ 2 \end{smallmatrix} \big| \begin{smallmatrix} 1 \\ 1 \end{smallmatrix}\right]_{\text{\tiny 6}}$, & {\footnotesize $\ldots$}, & \end{tabular} \\[4pt] \hline
\begin{tabular}{@{}c@{}}$\CP^1 \times {\cal B}_2$ \\ (${\cal B}_2$ almost \\ del Pezzo) \end{tabular} & 15,879,166 & \begin{tabular}{@{}c@{}}$\{8, 10, 12,$ \\ $14, 16, 18,$ \\ $20, 24\}$\end{tabular} & $\CP^1 \times\begin{dcases} \\ \\ \\ \\ \\ \end{dcases}$\hspace{-0.1cm}\begin{tabular}{@{}c@{}c@{}c@{}c@{}}
$\left[\begin{smallmatrix} 1 \\ 2 \end{smallmatrix} \big| \begin{smallmatrix} 1 \\ 1 \end{smallmatrix}\right]_{\text{\tiny 4}}$, & $\left[\begin{smallmatrix} 1 \\ 1 \\ 2 \end{smallmatrix} \Big| \begin{smallmatrix} 0 & 1 \\ 1 & 0 \\ 1 & 1 \end{smallmatrix}\right]_{\text{\tiny 5}}$, & $\left[\begin{smallmatrix} 1 \\ 1 \\ 1 \end{smallmatrix} \Big| \begin{smallmatrix} 1 \\ 1 \\ 1 \end{smallmatrix}\right]_{\text{\tiny 6}}$, & $\left[\begin{smallmatrix} 1 \\ 2 \end{smallmatrix} \big| \begin{smallmatrix} 1 \\ 2 \end{smallmatrix}\right]_{\text{\tiny 7}}$, \\ 
{\footnotesize$[4|2\;\,2]_{_{8}}$}, & {\footnotesize$[3|3]_{_{9}}$}, & $\left[\begin{smallmatrix} 1 \\ 2 \end{smallmatrix} \big| \begin{smallmatrix} 2 \\ 2 \end{smallmatrix}\right]_{\text{\tiny 10}}$, & $\left[\begin{smallmatrix} 1 \\ 2 \end{smallmatrix} \big| \begin{smallmatrix} 1 \\ 3 \end{smallmatrix}\right]_{\text{\tiny 12}}$, \\ 
$\left[\begin{smallmatrix} 2 \\ 2 \end{smallmatrix} \big| \begin{smallmatrix} 1 & 1 \\ 1 & 1 \end{smallmatrix}\right]_{\text{\tiny 6}}$, & $\left[\begin{smallmatrix} 1 \\ 1 \\ 1 \end{smallmatrix} \Big| \begin{smallmatrix} 1 \\ 1 \\ 2 \end{smallmatrix}\right]_{\text{\tiny 8}}$, & $\left[\begin{smallmatrix} 1 \\ 1 \\ 2 \end{smallmatrix} \Big| \begin{smallmatrix} 0 & 1 \\ 1 & 0 \\ 1 & 2 \end{smallmatrix}\right]_{\text{\tiny 8}}$, & $\left[\begin{smallmatrix} 1 \\ 3 \end{smallmatrix} \big| \begin{smallmatrix} 0 & 1 \\ 3 & 1 \end{smallmatrix}\right]_{\text{\tiny 12}}$, \\ 
$\left[\begin{smallmatrix} 1 \\ 4 \end{smallmatrix} \big| \begin{smallmatrix} 0 & 0 & 1 \\ 2 & 2 & 1 \end{smallmatrix}\right]_{\text{\tiny 12}}$, & $\left[\begin{smallmatrix} 1 \\ 1 \\ 1 \end{smallmatrix} \Big| \begin{smallmatrix} 1 \\ 2 \\ 2 \end{smallmatrix}\right]_{\text{\tiny 12}}$, & $\left[\begin{smallmatrix} 1 \\ 1 \\ 2 \end{smallmatrix} \Big| \begin{smallmatrix} 0 & 1 \\ 2 & 0 \\ 2 & 1 \end{smallmatrix}\right]_{\text{\tiny 12}}$
\end{tabular}\hspace{-0.1cm}$\begin{rcases} \\ \\ \\ \\ \\ \end{rcases}$ \\[4pt]
\bottomrule
\end{tabular}
\caption[Classification of base manifolds]{Classification of base manifolds that occur in our list. The first column lists the different types of three-fold bases. By ${\cal B}_3$ (${\cal B}_2$) we denote almost fano (almost del Pezzo) complete intersections in products of projective spaces, that is three-(two-)fold configurations whose anticanonical bundle is almost-ample. In contrast, ${\cal B}_3^\prime$ denote fano complete intersections in products of projective spaces. Their anticanonical bundle is ample. The subscripts on the configuration matrices denote the Euler characteristics. The second column counts how many times the types of base manifolds occur in the list of fibrations. In the third column, we list all of the different values for the Euler characteristic $\chi$ that occur. The last column contains example configurations and for the case of $\CP^1 \times {\cal B}_2$ this list is actually complete.\footnotemark}
\tablabel{bases}
\end{table}
We remark that bases of the form $(\CP^1)^2 \times {\cal B}_1$ and $\CP^2 \times {\cal B}_1$, where ${\cal B}_1$ is an almost ample complete intersection $1$-fold, such as $[2|2]$, $\left[\begin{smallmatrix} 1 \\ 1 \end{smallmatrix} \big| \begin{smallmatrix} 1 \\ 1 \end{smallmatrix}\right]$ or $\left[\begin{smallmatrix} 1 \\ 1 \end{smallmatrix} \big| \begin{smallmatrix} 1 \\ 2 \end{smallmatrix}\right]$, do not occur in the classification of base manifolds. This is a consequence of the redundancy removal (more precisely, the modding out by ineffective splittings and identities) that was employed in the compilation of the CICY 4-fold list~\cite{Gray:2013mja}, since the ${\cal B}_1$ merely describe different embeddings of $\CP^1$~\cite{Candelas:1987kf}. These cases are thus already captured by $(\CP^1)^3$ and $\CP^1 \times \CP^2$.

\vspace{0.1cm}

Of the 50,114,908 OEFs in our data set 26,088,498 satisfy the necessary condition for admitting a section which is a generic element of a favourable divisor class as described in \secref{sec}.
%
% see Table 1 (\tablabel{bases})
\footnotetext{Moreover, the topological types of almost del Pezzo surfaces are classified by their Euler characteristic, except for the case $\chi({\cal B}_2) = 4$~\cite{Hubsch:1992nu}. Hence, the configurations in the last two rows in the list of $\CP^1 \times {\cal B}_2$ configurations are equivalent to the respective configurations in the first two rows that have the same Euler characteristics.}
Restricting ourselves to the 49,466,248 elliptic fibrations which correspond to CICY four-folds which are not direct products, we find 25,999,860 examples which obey the conditions. \Figref{sec} shows the multiplicity of configuration matrices (omitting the direct products) which admit a given number of fibrations satisfying this necessary condition. The largest number of fibrations of a single configuration matrix potentially admitting a generic favourable section is 312. It should be noted that, because of the form of condition~\eqref*{secc}, any fibration which satisfies these conditions will admit multiple divisor classes which are suitable. This would correspond to, potentially, multiple sections (as opposed to a multi-section which all of the fibrations admit).
\begin{figure}[!t]\centering
\includegraphics[width=0.80\textwidth]{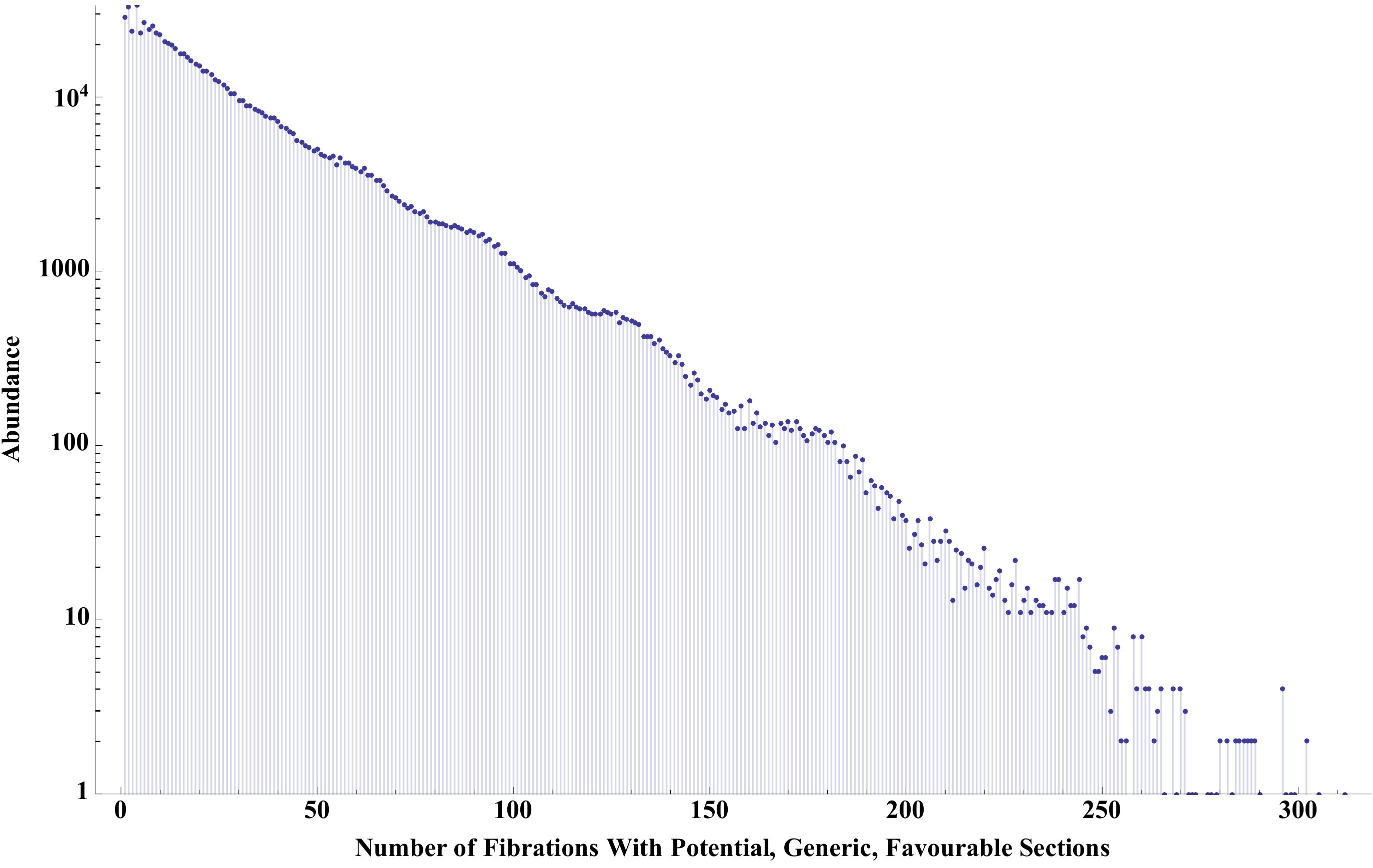}
\caption{Distribution of the multiplicity of configuration matrices (excluding product manifolds) admitting a given number of OEFs which satisfy the necessary conditions to admit a generic favourable section, as described in \secref{sec}.}
\figlabel{sec}
\end{figure}
%

%%%%%%%%%%%%%%%%%%%%%%%%%%%%%%%%%%%%%%%%%%%%%%%%%%%%%%%%%%%%%%%%%%%%%%%%%%%%%%%%%%%%%%%%%%%%%%%%
\section*{Acknowledgements}
%%%%%%%%%%%%%%%%%%%%%%%%%%%%%%%%%%%%%%%%%%%%%%%%%%%%%%%%%%%%%%%%%%%%%%%%%%%%%%%%%%%%%%%%%%%%%%%%

We would like to thank Lara Anderson, Ron Donagi and Martijn Wijnholt for useful discussions. A.~L.~is partially supported by the EPSRC network grant EP/l02784X/1. The results presented here were (partially) carried out on the cluster system at the Leibniz University of Hanover, Germany.

%%%%%%%%%%%%%%%%%%%%%%%%%%%%%%%%%%%%%%%%%%%%%%%%%%
\section*{Appendix}
\appendix
\section{Proof of efficient permutation redundancy removal algorithm}\applabel{perms}

In the compilation of the CICY four-fold list, it is important to know whether two matrices are related by a permutation of rows or columns, since in that case they correspond to a redundant description of the same underlying manifold. It is possible to partially remove this redundancy by imposing a lexicographic order on the rows and columns~\cite{Candelas:1987kf}. There remains, however, a residual redundancy and one may attempt to remove it by generating in a ``brute force'' way all row and column permutations of a matrix and comparing with the candidate equivalent configuration. This procedure eventually becomes unfeasible for the larger CICY configuration matrices in our list since the number of permutations grows exponentially with matrix size. 

In ref.~\cite{Gray:2013mja}, we thus developed a more efficient method of deciding if two matrices are related by permutation. The full details of this method and, in particular, the proof that it is equivalent to the aforementioned ``brute force'' procedure were omitted then, but shall be given in this appendix, instead. We mention in passing that a similar method can also be used to remove redundancies in the set of possible elliptic fibrations of a configuration with the important restriction that the permutations may not mix the submatrices describing fibre and base (see \secref{fib_class}).

Consider two configuration matrices, ${ A}$ and ${ \tilde{A}}$, of size $n\times m$. We will restrict ourselves to the situation where $n \leq m$. If this is not the case one can simply transpose the configuration matrix and then repeat the procedure elaborated below.  For the purposes of this section, we define an equivalence relation between such matrices as follows.
\begin{equation}
{ A} \sim { \tilde{A}} : \Leftrightarrow \exists \; \text{two permutation matrices} \; { P}\in  O(n) \;, { {\cal Q}} \in O(m) \; \text{such that} \; { A} = { P}^T { \tilde{A}} { {\cal Q}}
\end{equation}

Our goal is to find a necessary and sufficient criterion for ${ A}\sim { \tilde{A}}$ which is computationally efficient to implement. We start with a few basic facts about the singular value decomposition of an $n \times m$ matrix $A$. There exist orthogonal matrices ${ R} \in O(n)$ and ${ S} \in O(m)$ such that
\begin{equation}\eqlabel{singval}
 { R}^T { A} { S}=\left(\begin{array}{cc} { D}&0\end{array} \right)\; ,\quad { D}=\text{diag}(a_1,\ldots ,a_n)\; .
\end{equation} 
Multiplying this equation with its transpose gives
\begin{equation}
 { R}^T{ A} { A}^T { R}={ D}^2=\text{diag}(a_1^2,\ldots ,a_n^2)\; .
\end{equation} 
This shows that the columns of ${ R}$ consist of the eigenvectors of ${ A}{ A}^T$. Let us assume that the spectrum $\{a_i^2\}$ is non-degenerate. In this case the (normalized) eigenvectors in ${ R}$ are uniquely determined up to permutations and a sign ambiguity for each of them. 
Ordering the eigenvalues $a_i^2$ (say with increasing size) removes the permutation ambiguity. Further, let us assume that $\sum_{i=1}^nR_{ij}\neq 0$, that is, all eigenvectors are such that their component sum is different from zero. In this case, the sign ambiguity can be fixed by demanding that 
\begin{equation}\eqlabel{pos}
 \sum_{i=1}^nR_{ij}> 0
\end{equation}
 for all $j=1,\ldots ,n$. So, in summary, provided the eigenvalues of ${ A}{ A}^T$ are non-degenerate and ordered and the conditions~\eqref*{pos} are satisfied the diagonalizing matrix ${ R}\in O(n)$ in the singular value decomposition~\eqref*{singval} is unique. With these preliminaries complete, we are now in a position to formulate the following claim.\\[0.3cm]
{\bf Claim:} Let ${ A},{ \tilde{A}}$ be two $n\times m$ (where $n\leq m$) matrices with the same, non-degenerate spectrum of eigenvalues for ${ A}{ A}^T$ and ${ \tilde{A}}{ \tilde{A}}^T$. Further, let ${ R},{ \tilde{R}}\in O(n)$ be diagonalizing matrices, that is ${ R}^T{ A}{ A}^T{ R}={ \tilde{R}}^T{ \tilde{A}}{ \tilde{A}}^T{ \tilde{R}}={ D}^2=\text{diag}(a_1^2,\ldots ,a_n^2)$, which both satisfy condition~\eqref*{pos}. Then it follows that
\begin{align*}
 { A}\sim{ \tilde{A}}\Longleftrightarrow &({ P}:={ \tilde{R}}{ R}^T \text{ is a permutation and }\\ &\qquad\qquad{ A},{ A'}:={ P}^T{ \tilde{A}}\text{ have the same sets of column vectors.})
\end{align*}
{\bf Proof:} \vspace{-0.4cm}
\paragraph{``${\boldsymbol \Longrightarrow}$"{\bf :}} Let ${ A}\sim{ \tilde{A}}$ so that ${ A}={ P}^T{ \tilde{A}} { {\cal Q}}$ for two permutation matrices ${ P},{ {\cal Q}}$. It follows that ${ A}{ A}^T={ P}^T{ \tilde{A}}{ \tilde{A}}^T{ P}$, so that ${ D}^2={ R}^T{ A}{ A}^T{ R}=({ PR})^T{ \tilde{A}}{ \tilde{A}}^T({ PR})$. On the other hand ${ D}^2={\tilde{R}}^T{ \tilde{A}}{ \tilde{A}}^T{ \tilde{R}}$, so that ${ \tilde{A}\tilde{A}}^T$ is diagonalized by ${ PR}$ and ${ \tilde{R}}$. Both ${ \tilde{R}}$ and ${ R}$ satisfy condition~\eqref*{pos} by assumption. It follows that also ${ PR}$ does because multiplication with ${ P}$ from the left corresponds to row permutation which leaves the sums~\eqref*{pos} unchanged. The above uniqueness statement then implies that ${ \tilde{R}}={ PR}$ so that ${ P}={ \tilde{R}R}^T$ is indeed a permutation. Then ${ A}={ A'}{ {\cal Q}}$, for a permutation ${ {\cal Q}}$, so that ${ A}$ and ${ A'}$ have the same column vectors, possibly with different ordering. \vspace{-0.3cm}
\paragraph{``${\boldsymbol \Longleftarrow}$"{\bf :}} Assume that ${ P}={ \tilde{R}R}^T$ is a permutation and ${ A,A'}={ P}^T{ \tilde{A}}$ have the same column vector sets. From ${ R}^T{ AS}={ \tilde{R}}^T{ \tilde{A}\tilde{S}}$ it follows that ${ A}={ P}^T{ \tilde{A}}{ {\cal Q}}={ A}'{ {\cal Q}}$, where ${ {\cal Q}}={ \tilde{S}S}^T\in O(n)$. Since ${ A}$ and ${ A'}$ only differ by a permutation of columns we can choose ${ {\cal Q}}$ to be a permutation matrix, so that ${ A}={ P}^T{ \tilde{A}}{ {\cal Q}}$ with two permutation matrices ${ P},{ {\cal Q}}$ and, hence, ${ A}\sim{ \tilde{A}}$. $\Box$\\[0.3cm]
So, in practice, we first check that the spectrum of ${ AA}^T$, ${ \tilde{A}\tilde{A}}^T$ is identical (if it is not the configurations are of course inequivalent) and non-degenerate and then find the corresponding diagonalizing matrices ${ R},{ \tilde{R}}\in O(n)$ for the same ordering of eigenvalues and ensure they both satisfy condition~\eqref*{pos}.\footnote{If the spectrum happens to be degenerate, or the sum in~\eqref*{pos} vanishes, then we can modify the configuration matrices ${ A}$ and $\tilde{ A}$ in a way that does not affect equivalence but may change the spectrum or positivity properties. For example, we can substitute the occurrence of a given number everywhere in the matrix with a different value. If, in a given case, such procedures can not change the eigenvalues of the matrices ${ A} { A}^T$ and ${ \tilde{A}}{ \tilde{A}}^T$ then the brute force method described earlier must be employed. If condition~\eqref*{pos} cannot be satisfied because one or more eigenvectors have components which sum to zero, one can compute ${ R}$ and ${ \tilde{R}}$ for all possible sign choices for those eigenvectors and then check if the criterion is satisfied for at least one such choice.} Then we compute ${ P}={ \tilde{R}R}^T$ and check if it is a permutation matrix. If it is not, the configurations are inequivalent. If it is we compute ${ A'}={ P}^T{ \tilde{A}}$ and check if is has the same column vector set as ${ A}$. If it does the two configurations are equivalent, otherwise they are not.

\section{Data Format}\applabel{dataformat}

In this appendix, we describe the format in which the data that was computed in this project is stored and made available. The complete data is contained in two data sets which can be downloaded from~\cite{cicylist4} in compressed form. The first data set includes the configuration matrices and the topological invariants associated to them. It is stored as a \emph{Mathematica} list \verb+{configuration1, configuration2, configuration3, ...}+ and can be loaded in \emph{Mathematica} using the \verb+ReadList+ command. Each entry has the following structure
\begin{equation*}
 \verb+{Id, +m\verb+, +K\verb+, +{\bf q}\verb+, +\chi\verb+, IsProduct, +h^{1,1}\verb+, +h^{2,1}\verb+, +h^{3,1}\verb+, +h^{2,2}\verb+, Favour}+ ,
\end{equation*}
where the unique identifier \verb+Id+ is a positive integer indicating the position of the entry in the full list, $m$ is the number of ambient space projective factors, $K$ the number of defining polynomials, ${\bf q}$ the configuration matrix,\footnote{The configuration matrix is stored without the column vector {\bf n}, which contains the dimensions of the ambient space projective factors. It is straightforward to reconstruct {\bf n} using the Calabi-Yau condition~\eqref*{c1zero}.} $\chi$ the Euler characteristic and $h^{p,q}$ the Hodge numbers. \verb+IsProduct+ is \verb+true+ if the configuration matrix is block diagonal thus representing a product manifold and \verb+false+ otherwise. \verb+Favour+ is \verb+true+ if the configuration matrix is favourable in the sense explained in \secref{hodgecalc} and \verb+false+ otherwise.

The second data set contains the fibration structure of the manifolds. It is also stored as a \emph{Mathematica} list and split across several files of manageable size. Each entry is of the form
\begin{equation*}
 \verb+{Id, Fibs}+ .
\end{equation*}
Here, \verb+Id+ is the same unique identifier that was introduced above and configurations that do not have OEFs are omitted from the list. \verb+Fibs+ represents a list of lists where each entry has the form
\begin{equation*}
 \verb+{FibEntries, {SectionCond, S}}+ ,
\end{equation*}
where \verb+FibEntries+ is a list of two lists. The first list are the rows and the second list the columns corresponding to the sub-part, $[{\cal A}_1| {\cal F}]$, of the configuration matrix which describe the fibre. \verb+SectionCond+ is \verb+true+ if the section condition~\eqref*{secc} is satisfied. In that case, \verb+S+ contains the components $a^r$ of the two-form $S=a^r J_r$ as defined\footnote{It should be noted that the labelling of the $J_r$ in $S$ is with respect to the OEF form~\eqref*{mrfib}, which does not necessarily have the same ordering of rows and columns as the original configuration matrix.} in \secref{sec}. On the other hand, if the section condition~\eqref*{secc} cannot be satisfied, \verb+SectionCond+ is \verb+false+ and \verb+S+ is the empty list \verb+{}+.

The data format is best illustrated by means of an example:
\begin{center}
\begin{minipage}{11.8cm}
 \verb+{1595, 4, 3, {{0, 1, 1}, {2, 0, 0}, {2, 0, 1}, {1, 1, 2}},+\\
 \verb+ 648, False, 4, 0, 96, 444, True}+ .
\end{minipage}
\end{center}
This is the 1595th entry in the CICY four-fold list represented by the configuration matrix
\begin{equation*}
 \left[\begin{array}{c|ccc} 1 & 0 & 1 & 1 \\ 1 & 2 & 0 & 0 \\ 2 & 2 & 0 & 1 \\ 3 & 1 & 1 & 2 \end{array}\right] \; .
\end{equation*}
It is not block-diagonal, but favourable and has topological data $\chi = 648$, $h^{1,1}=4$, $h^{2,1}=0$, $h^{3,1}=96$ and $h^{2,2}=444$.
The corresponding entry in the fibration data set reads
\begin{center}
\begin{minipage}{10.7cm}
 \verb+{1595, {{{{1, 4}, {1, 2, 3}}, {True, {2, -1, 0, 0}}},+\\
 \verb+        {{{2, 3}, {1, 3}}, {False, {}}},+\\
 \verb+        {{{2, 4}, {1, 2, 3}}, {False, {}}},+\\
 \verb+        {{{1, 2, 3}, {1, 2, 3}}, {False, {}}}}}+ .
\end{minipage}
\end{center}
Thus, there are four OEFs, namely $\left[\begin{smallmatrix} 1 \\ 3 \end{smallmatrix} \big| \begin{smallmatrix} 0 & 1 & 1 \\ 1 & 1 & 2 \end{smallmatrix}\right]$ fibered over $\CP^1 \times \CP^2$, $\left[\begin{smallmatrix} 1 \\ 2 \end{smallmatrix} \big| \begin{smallmatrix} 2 & 0 \\ 2 & 1 \end{smallmatrix}\right]$ fibered over $\left[\begin{smallmatrix} 1 \\ 3 \end{smallmatrix} \big| \begin{smallmatrix} 1 \\ 1 \end{smallmatrix}\right]$, $\left[\begin{smallmatrix} 1 \\ 3 \end{smallmatrix} \big| \begin{smallmatrix} 2 & 0 & 0 \\ 1 & 1 & 2 \end{smallmatrix}\right]$ fibered over $\CP^1 \times \CP^2$ and $\left[\begin{smallmatrix} 1 \\ 1 \\ 2 \end{smallmatrix} \Big| \begin{smallmatrix} 0 & 1 & 1 \\ 2 & 0 & 0 \\ 2 & 0 & 1 \end{smallmatrix}\right]$ fibered over $\CP^3$. Only for the first of the four OEFs, the section condition~\eqref*{secc} is satisfied. With respect to the OEF form of the configuration matrix
\begin{equation*}
 \left[\begin{array}{c|ccc} 1 & 0 & 1 & 1 \\ 3 & 1 & 1 & 2 \\ 1 & 2 & 0 & 0 \\ 2 & 2 & 0 & 1 \end{array}\right] \; ,
\end{equation*}
the corresponding two-form $S$ is then given by $S = 2 J_1 - J_2$.

%%%%%%%%%%%%%%%%%%%%%%%%%%%%%%%%%%%%%%%%%%%%%%%%%%%%%%%%%%%%%%%%%%%%%%%%%%%%%%%%%%%%%%%%%%%%%%%%

\end{document}